\documentclass[aps,prb,twocolumn,superscriptaddress]{revtex4-1}

\usepackage{amsmath,amssymb} 
\usepackage{physics} 
\usepackage{graphicx}
\usepackage{color}
\usepackage{dcolumn}
\usepackage{bm}
\usepackage[mathlines]{lineno}
\newcommand\numberthis{\addtocounter{equation}{1}\tag{\theequation}}

\begin{document}


\title{Angle-resolved photoemission intensity for multi-orbital bands: Complex interplay between the self-energy matrix and the optical matrix elements}

\author{Yau Chuen Yam}
\author{Mona Berciu}
\author{George A. Sawatzky}
\affiliation{\!Department \!of \!Physics and Astronomy, \!University
	of\!  British Columbia, \!Vancouver, British \!Columbia,\! Canada,\!
	V6T \!1Z1}
\affiliation{Quantum Matter \!Institute, \!University
	of British Columbia, \!Vancouver, British \!Columbia, \!Canada,
	\!V6T \!1Z4}

\date{\today}

\begin{abstract}
  We use a simple one-dimensional two-band model with electron-phonon coupling to illustrate some of the complications that arise in multi-band systems when trying to extract a self-energy using the typical approach used for single-band systems when analyzing  angle-resolved photoemission spectroscopy (ARPES) data. The underlying reason is that in multi-band models the self-energy is a matrix, not a scalar,  and the result obtained from the ARPES analysis is a complicated function of all these self-energy matrix elements, weighted by  different  dipole matrix elements of the relevant Wannier orbitals. We contrast the results for Holstein and Peierls electron-phonon couplings to further illustrate differences between models with a local versus non-local self-energy matrix.
\end{abstract}

\pacs{Valid PACS appear here}
\maketitle

\section{\label{sec:intro}Introduction}

The selfenergy $\Sigma({\bf k},\omega)$ is an essential quantity needed for characterizing quasiparticle properties in an interacting system (in this work, we assume that a quasiparticle description is valid at low energies). In one-band models it determines the single-particle propagator:
\begin{equation}
\label{e1.1}
G({\bf k},z) = \frac{1}{z - \epsilon_{\bf k}-\Sigma({\bf k},z)}
\end{equation}
and differentiates it from the non-interacting propagator 
\begin{equation}
\label{e2}
G_0({\bf k},z) = \frac{1}{z - \epsilon_{\bf k}}.
\end{equation}
Here, $\epsilon_{\bf k}$ is  the bare band  dispersion as a function of the crystal momentum ${\bf k}$,  $z=\omega \pm i\eta$ is the energy with a small artificial broadening $\eta\rightarrow 0$, and the sign $\pm$ corresponds to electron addition/removal propagators corresponding to momenta outside/inside the Fermi sea.
Knowledge of the selfenergy allows us to find the dispersion of the dressed band $E_{\bf k}$ from the poles of $G({\bf k},\omega)$:
\begin{equation}
\label{e3}
E_{\bf k} =  \epsilon_{\bf k}+\Sigma({\bf k},E_{\bf k})
\end{equation}
Generically, the stronger the interactions, the larger the self-energy and hence the bigger the difference between the dispersion of  quasiparticles and bare particles. 

From the theory side, there are a few ways to calculate the self-energy such as  diagrammatics and variational approximations, although more methods are needed. From the experimental side, the most direct access to this quantity is through angle-resolved photoemission spectroscopy (ARPES), which measures an intensity\cite{Dam2003} 
\begin{equation}
\label{e3.5}
I_{exp}({\bf K},\omega) \propto |M({\bf K}, {\bf k})|^2 A({\bf k}, \omega) f(\omega)
\end{equation}
Here, $f(\omega)$ is the Fermi-Dirac distribution, $M({\bf K}, {\bf k})$ is a dipole matrix element that depends both on the momentum ${\bf K}$ of the photoelectron and on the crystal momentum ${\bf k}$ of the quasiparticle involved in the photoemission process (this quantity and its relation to  ${\bf k}$,  ${\bf K}$ are discussed below), and the electron-removal spectral weight is defined as  $A({\bf k},\omega) = \frac{1}{ \pi} \mbox{Im} G({\bf k},\omega)$, so that
\begin{equation}
\label{e4}
A({\bf k},\omega) \propto \frac{\Sigma"({\bf k},\omega)}{(\omega - \epsilon_{\bf k} -\Sigma'({\bf k},\omega))^2 + (\Sigma"({\bf k},\omega))^2}
\end{equation}
where $\Sigma', \Sigma"$ are the real and imaginary parts of the self-energy, respectively. If the dipole matrix element has a weak momentum dependence, one can attribute features in the ARPES intensity to features in the spectral weight and therefore extract the dispersion $E_{\bf k} =  \epsilon_{\bf k}+\Sigma'({\bf k},E_{\bf k})$ from the position of the lowest binding energy peak, and $\Sigma"({\bf k},E_{\bf k})$ from the peak broadening. If $\epsilon_{\bf k}$ is available from ab-initio calculations, then one can find the selfenergy $\Sigma(E_{\bf k}, {\bf k})$ and compare it to theoretical predictions.

While this type of analysis is routinely performed for a large variety of materials, everything stated above requires, as a  minimal necessary condition, that there is a single band close to the Fermi energy, well separated in energy from all other bands. In fact, as we show below, even this is not sufficient to guarantee the validity of this kind of analysis. In addition, it is necessary to also have vanishing interband matrix elements of the interaction and/or coupling responsible for the quasiparticle renormalization.

Most quantum materials, however, either (i) have multiple bands close to the Fermi energy, or (ii) if a single band crosses $E_F$, its underlying Wannier orbitals are linear combinations of multiple atomic orbitals, with coefficients strongly dependent on ${\bf k}$, or (iii) even if a band originating from a single type of atomic orbital is well separated from the other bands in an ab-initio study, it is still possible that strong correlations and/or strong electron-phonon coupling could lead to significant shifts of spectral weight between several bands. Any of these scenarios would render the validity of an analysis that ignores the presence of other bands highly suspect.

At first sight, generalizing the analysis to a multi-band system appears to be trivial: first, the propagators and the self-energy now become matrices, with a dimension set by the number $n$ of mixing bands. Second, the link between these matrices is:
\begin{equation}
\label{e5}
[G({\bf k}, \omega)]^{-1} = [G_0({\bf k}, \omega)]^{-1}- [\Sigma({\bf k}, \omega)]
\end{equation}
where $[\dots]^{-1}$ denotes matrix inversion.

Equation (\ref{e5}) illustrates the main origin of complications in the multi-band case. One could, in principle, choose a basis in which the matrix  $[G({\bf k}, \omega)]$ is diagonal, so that one can analyze the different renormalized bands independently (this basis would be different for each ${\bf k}$ value, though). However,   $[G_0({\bf k}, \omega)]$ and $[\Sigma({\bf k}, \omega)]$ are {\em not} diagonal in this same basis (if they were, then the model would reduce trivially to $n$ independent models). As a result, the attempt to treat each individual dressed band as if it were described by its own single-band propagator with its own self-energy like in Eq. (\ref{e1.1}) will fail because it would only identify  $n$ 'self-energies', whereas the self-energy matrix has $n\times n$ components. In other words, it is not possible to extract the full many-band self-energy from analyzing only peak locations and broadenings associated with individual dressed bands.

These bad news, insofar as analysing ARPES  in multi-band systems is concerned, are further compounded by the fact that $I_{exp}({\bf K},\omega)$ now involves a combination of various matrix elements of $[G({\bf k}, \omega)]$ weighted by various matrix elements $[M({\bf K}, {\bf k})]$ associated with the different atomic orbitals defining the model (see below). The momentum dependence of the latter can vary significantly for different orbitals, and disentangling these contributions from those arising from various $[G({\bf k}, \omega)]$ becomes very difficult, if not outright impossible. 

In this work we illustrate some of these issues using the simplest one-dimensional model with only two bands, and a self-energy arising from electron-phonon coupling (EPC). The latter is a convenient choice because it allows us to contrast the differences of the predicted $I_{exp}({\bf K},\omega)$  for a model with Peierls EPC, which produces a self-energy $[\Sigma_P({\bf k}, \omega)]$ with nontrivial ${\bf k}$ dependence, versus a model with Holstein EPC which produces a self-energy $[\Sigma_H(\omega)]$ that is essentially local. The physical origin of these two types of couplings is discussed below. 

The work is organized as follows:  Section II describes the models and the methods we use to calculate their self-energies. Section III presents the results, and Section IV contains their summary and discussion.

\section{\label{method}Model and methods}

\begin{figure}[t]
	\centering
	\includegraphics[width=0.55\columnwidth]{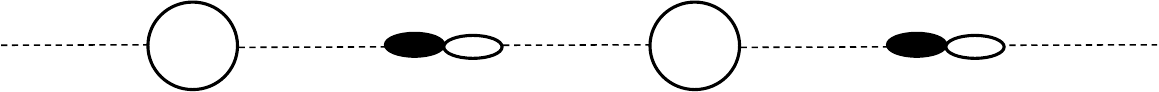}
        \vspace{5mm}
        
        \includegraphics[width=0.55\columnwidth]{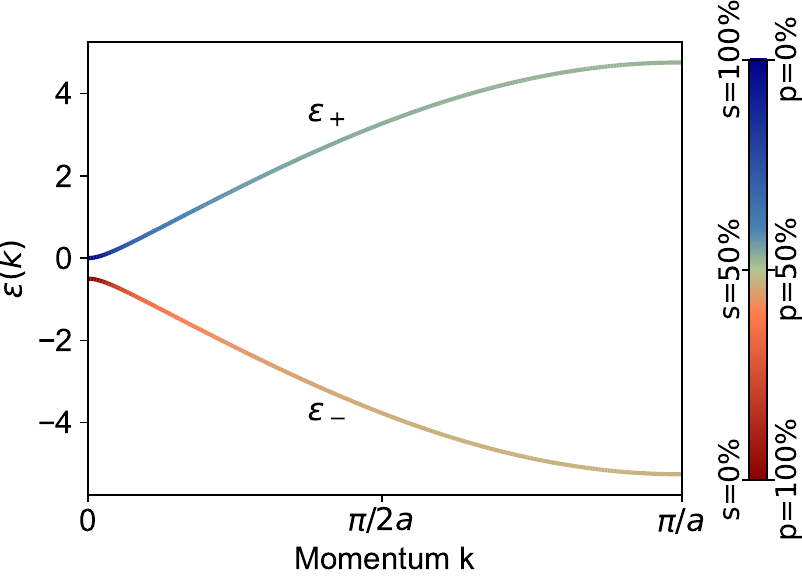}
	\caption{\label{1D_chain} (a) Chain with alternating $s$ and $p_x$ valence orbitals. The atoms hosting the $s/p$ orbitals are assumed to be heavy/light and thus immobile/oscillating; (b)  Dispersion of the two bare bands in the positive half of the Brillouin zone, for $\Delta =0.5, t=2.5$. For these values the GS is at $\epsilon_{GS}\approx -5.256$, $k_{GS}=\pi$. The colour indicates the $p$ (red) vs. $s$ (blue) character of the eigenfunctions. }
\end{figure}

As already mentioned, here we study the simplest  possible  one-dimensional two-band model in order to illustrate  non-trivial behaviours that we believe to be generic for all  multi-band systems. Specifically, we  assume  alternating $s$ and $p$ valence orbitals, as sketched in Fig. \ref{1D_chain}(a). Unit cell $i$ contains the orbitals $s$ located at $R_i=ia$ and $p$ located at $R_i+a/2$. Hereafter we set $a=1$. Defining  $s_{i\sigma}^{\dagger}$ and $p^{\dagger}_{i\sigma}$ as the corresponding creation  operators for holes in these orbitals, the minimal electronic Hamiltonian  describing this system is:
\begin{equation}
{\cal H}_{0,\rm{e}} = -t\sum_{i\sigma} [s_{i\sigma}^{\dagger}(p_{i\sigma}-p_{i-1,\sigma})+\text{h.c.}]-\Delta\sum_{i,\sigma} p^{\dagger}_{i\sigma} p_{i\sigma}
\label{e1}
\end{equation}
Here, $t>0$ is magnitude of the hopping between neighbor $s$ and $p$ orbitals when they are at their equilibrium positions, while $\Delta$ characterizes the charge transfer energy. We ignore correlations because we will investigate a {\em single hole} in an otherwise full band. (Studying instead a single electron in an otherwise empty band would not lead to any qualitative difference for the results of interest to us). The spin of this hole is irrelevant, for the same reason, so from now on we drop the spin index.

This bare Hamiltonian can be diagonalized straightforwardly. The energies of its two  bands are
\begin{equation}
\epsilon_{\pm}(k) = {1\over 2} \left(-\Delta \pm \sqrt{\Delta^2 + 16 t^2 \sin^2 {k\over 2} }\right) 
\label{disp0}
\end{equation}
They are shown  in Fig. \ref{1D_chain}(b) for the values $\Delta=0.5, t=0.25$ that we will use throughout this work. As expected for a hole, the GS is at $k_{GS}=\pi$ and its energy is $\epsilon_{GS}\approx -5.256$, and for these parameters has nearly equal $s$ and $p$ character (it is precisely equal if $\Delta =0$). This significant mix of orbital character persists at most momenta except at $k=0$, where it is forbidden by symmetry. Specifically, at $k=0$ we find that $\epsilon_-(0)=-\Delta$ is of pure $p$ character, while $\epsilon_+(0)=0$ is of pure $s$ character.

To introduce an optical phonon mode, we assume that the atoms hosting the $p$ orbitals are much lighter than those hosting the $s$ orbitals, as is the case in perovskites like BaBiO$_3$, where the relevant orbitals are $6s$ for the heavy Ba cation and $2p$ for the light O anion. As a result, to first order the motion of the latter can be ignored, and the motion of the former can be described as longitudinal independent harmonic oscillators in the potential well created by the immobile fixed heavy neighbors. This simplifies the longitudinal optical phonons to an Einstein model of energy $\Omega$ (we set $\hbar=1$):
\begin{equation}
{\cal H}_{0,\rm{ph}} = \Omega\sum_i b^{\dagger}_i b_i
\label{e1.2}
\end{equation}
where $b^{\dagger}_i, b_i$ are the phonon  creation and annihilation operators. The displacement out of equilibrium of site $p$ in the unit cell $i$ is, then: $\hat{u}_i =  (b^{\dagger}_i+ b_i)/\sqrt{2M\Omega}$, where $M$ is the mass of this atom. Of course, a more detailed modeling of the optical phonon is possible and will lead to a momentum-dependent dispersion $\Omega_q$. As we discuss in more detail below, this would only further amplify the momentum dependence of the self-energy matrix.

The motion of the $p$ orbitals influences the behavior of the hole through EPC. Roughly speaking, there are two main types of EPC, with different underlying origins. The first dominates when the system is rather covalent, {\em i.e.} when $|\Delta| < |t|$, and arises from the modulation of the $s$-$p$ hopping due to the motion of the $p$ orbitals. Assuming small displacements and using the linear approximation, this leads to the Peierls model
\begin{align*}
{\cal H}_{P}= {\cal H}_0+ g\sum_i [p^{\dagger}_i(s_i+s_{i+1})(b^{\dagger}_i+b_i)
+\text{h.c.}]
\numberthis
\label{Peierls}
\end{align*}
where $g>0$ characterizes the electron-phonon coupling, and  ${\cal H}_0= {\cal H}_{0,\rm{e}} +{\cal H}_{0,\rm{ph}}$. 

In the ionic limit ($|\Delta|\gg |t|$), on the other hand, the main source of EPC comes from the modulation of the Coulomb interaction (Madelung potential) between the hole and nearby ions, due to the lattice motion. If the hole is on an $s$ orbital, then anti-phase `breathing-mode' motion of the two nearby $p$ orbitals will strongly modulate its onsite energy. In contrast, if the hole is on a $p$ orbital, to first order its on-site energy is not affected by the motion of the ion it sits on. This is because any displacement brings the hole closer to one and further from the other of its neighbors, and the modulations of the corresponding Coulomb interactions cancel out to linear order in the displacement (we ignore quadratic and higher order EPC  in this work). Such a breathing-mode EPC is described by a Hamiltonian of the type ${\cal H}_{BM}={\cal H}_0+g\sum_i s^{\dagger}_i s_i (b^{\dagger}_i+b_i-b^{\dagger}_{i-1}+b_{i-1})$ .\cite{BLau2007} Instead, we choose to study the less appropriate but simpler Holstein model:
\begin{align*}
{\cal H}_{H}={\cal H}_0+g\sum_i p^{\dagger}_i p_i (b^{\dagger}_i+b_i)
\numberthis
\label{Holstein}
\end{align*}
with Holstein coupling at the $p$ sites. Given the symmetry of the system, such a coupling cannot arise if $b^{\dagger}_i, b_i$ describe the actual motion of the $p$ orbitals, as discussed above. In the Holstein spirit, \cite{HolsteinI}  we instead take them to now describe vibronic distortions of `polar molecules' whose valence state is described by the $p$ orbital.

We choose the Holstein EPC as our second option because, as we show below, its corresponding self-energy is nearly local and therefore allows us to understand the relevance of having a momentum independent self-energy (the Peierls EPC leads to a momentum dependent self-energy, as does the breathing-mode EPC). We emphasize, however, that the breathing-mode EPC can be studied with the method discussed below \cite{GGoodvinPRB2008} and the same is true for models with dual Peierls+Holstein \cite{DMarchandPRB2017} or Peierls+breathing mode EPC.\cite{HerreraPRL2013}  While the latter are physically more relevant for any system that is neither purely ionic or purely covalent, studying such dual models is not shedding any additional light on the issues we are investigating here.

For our purposes, the crucial difference between the Peierls and Holstein couplings is that in momentum space, the vertex of the former depends on the electron momentum $k$ ({\em i.e.} this is a so-called $g(k,q)$ coupling), whereas that of the latter does not (this is a so-called $g(q)$ model). Indeed, if we apply the Fourier transformation:
 \begin{align*}
c^\dagger_i=\sum_k\frac{e^{-ikR_i}}{\sqrt{N}}c^\dagger_k
  \numberthis
  \label{ops}
 \end{align*}
 where $c=s, p, b$, we find that:
 \begin{align*}
   {\cal H}_{P}= &{\cal H}_0+
\sum_{k,q}\frac{g}{\sqrt{N}}\Big[(1+e^{-i(k-q)})s^{\dagger}_{k-q}p_k\\
& + (1+e^{ik})p^{\dagger}_{k-q}s_k\Big](b^{\dagger}_{q}+b_{-q})
  \numberthis
  \label{H_Pei}
 \end{align*}
has explicit $k$ dependence in the EPC, whereas
\begin{align*}
{\cal H}_{H}	={\cal H}_0+ \sum_{k,q}\frac{g}{\sqrt{N}}
	p^\dagger_k p_{k+q}(b^\dagger_q+b_{-q})
	\numberthis\label{H_Hol}
\end{align*}
does not. Here, $N \rightarrow \infty$ is the number of unit cells.

Single polaron results for this Peierls model have already been studied in Ref. \onlinecite{Mol2016} in 1D; qualitatively similar polaron behavior was then demonstrated for its generalization to 2D Lieb and 3D perovskite-type lattices in Refs. \onlinecite{Mol2017} and \onlinecite{Yam2020}, respectively. One difference between the 1D system and those in higher dimension lies in the spatial separation of orbitals of the same kind, which is $a$ in the 1D system but  only $a/\sqrt{2}$ in 2D; as a result, some properties may be affected when the dimension is different. However, even in higher dimension we expect the $s-s$ and $p-p$ hopping to remain considerable smaller than the $s-p$ hopping, because the distance between adjacent $s$ and $p$ orbitals is still much smaller than that between orbitals of the same kind. 

As already mentioned, here we analyze the self-energy of these models in the single carrier (hole) sector. We therefore use the same approach described in Ref. \onlinecite{Mol2016}, namely the 
Momentum Average (MA) variational approximation \cite{Ber2006,Ber2007} which allows us to calculate the Green's function matrix:
\begin{align*}
G(k,\omega) = \Big(\begin{matrix}
G_{ss}(k,\omega) & G_{sp}(k,\omega)\\
G_{ps}(k,\omega) & G_{pp}(k,\omega)\\
\end{matrix}\Big)
	\numberthis\label{propm}
\end{align*}
where, for example,  
\begin{align*}
G_{sp}(k,\omega) = \langle0| s_k \hat{G}(\omega) p_k^\dagger|0\rangle
	\numberthis\label{prop}
\end{align*}
and $\hat{G}(\omega)= (\omega+i\eta-{\cal H})^{-1}$ is the rezolvent for the Hamiltonian of interest. We define the self-energy matrix elements in this basis as:
\begin{align*}
\Sigma_{\alpha\beta}(k,\omega) = [ G(k,\omega)]^{-1}_{\alpha\beta} - [ G_0(k,\omega)]^{-1}_{\alpha\beta} 
	\numberthis\label{self}
\end{align*}
where $\alpha,\beta=s,p$ and $G_0(k,\omega)$ is the Green's function matrix associated with the non-interacting Hamiltonian ${\cal H}_0$, in the same basis. 

For completeness, we mention that we also used Exact Diagonalization (ED) to verify that the polaron bands extracted from the poles of the MA $G(k,\omega)$ are in good agreement with those found by ED. We use MA simply because it gives easier access to the self-energy. We also note that previous work shows that MA tends to {\em understimate} the momentum-dependence of the self-energy,\cite{Ber2007} so a more accurate calculation should find even stronger momentum dependence of the self-energy than the one illustrated here.

\section{\label{sec:results}Results}

\subsection{Effects of the EPC on the Spectral Weights}

\begin{figure}
	\centering
	\includegraphics[width=1\columnwidth]{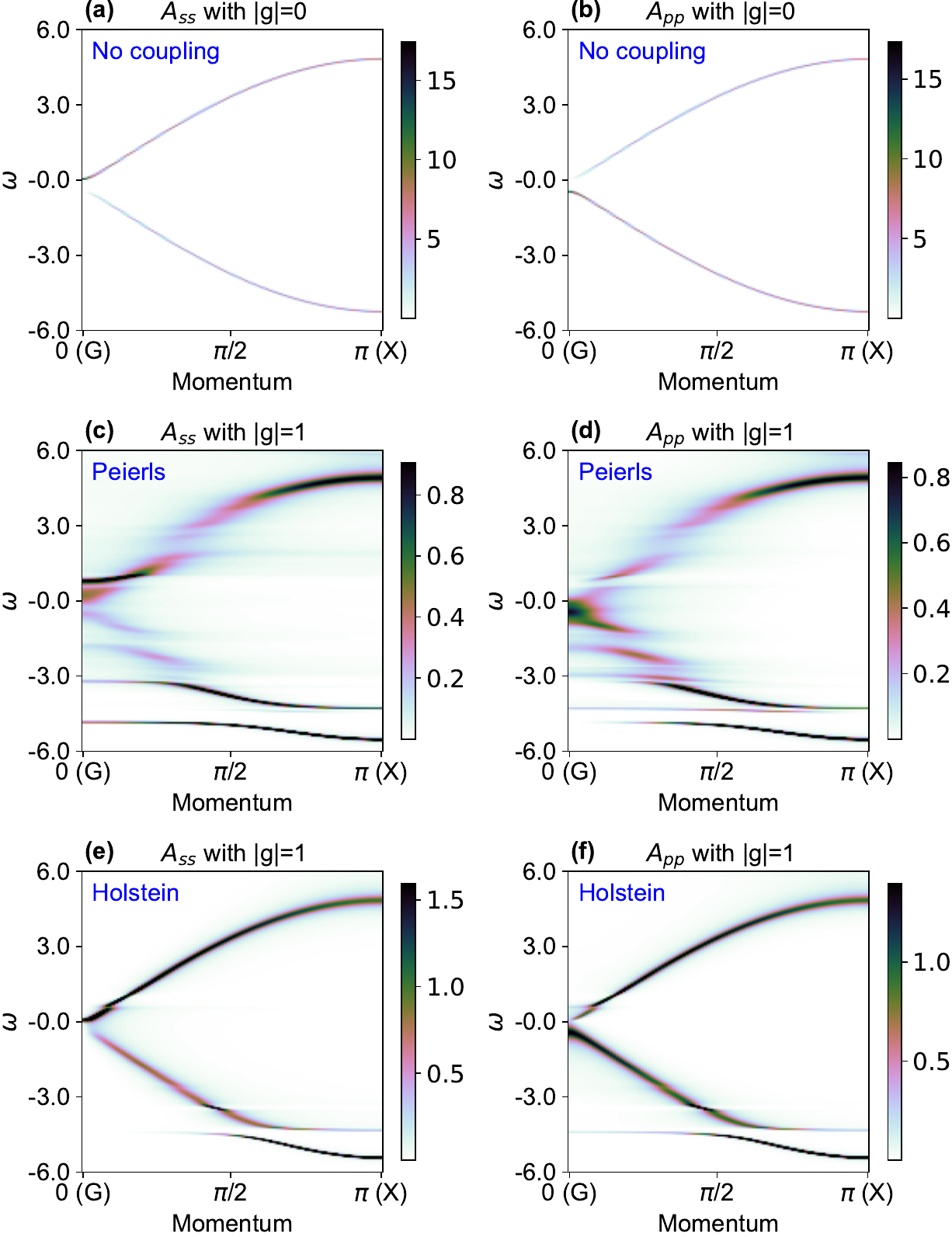}
	\caption{\label{sw_t_not0} Contour plots of $A_{ss}({\bf k}, \omega)$ (left panels) and $A_{pp}({\bf k}, \omega)$ (right panels) and (a-b) no electron phonon coupling, (c-d) Peierls coupling and (e-f) Holstein coupling. Other parameters are $t=2.5, \Omega=1, \Delta=0.5$. Peak broadenings are  $\eta=5\cdot 10^{-6}$ and $\eta=5\cdot 10^{-4}$  for the Peierls and Holstein EPC, respectively. }
\end{figure}

The effect of these EPC on the electronic spectra is revealed in  Fig. \ref{sw_t_not0}, where we plot the hyperbolic tangent of the spectral weights $A_{ss}(k,\omega) = - {1\over \pi} \mbox{Im} G_{ss}(k,\omega)$ (left column) and $A_{pp}(k,\omega) = - {1\over \pi} \mbox{Im} G_{pp}(k,\omega)$ (right column) for the uncoupled case (top row), and for the  Peierls and Holstein models (middle and bottom rows, respectively). We plot the energy range spanning both bare bands, together with the polaron eigenstate $|P;k\rangle$, which is visible as the lowest-energy feature in both   $A_{ss}(k,\omega), A_{pp}(k,\omega)$. The different associated weights reflect the overlaps $|\langle P;k | s^\dagger_{k}|0\rangle|^2$ and $|\langle P;k | p^\dagger_{k}|0\rangle|^2$, revealing the orbital character of the polaron eigenstate.

As already discussed, for the uncoupled case with $g=0$ (top row), the GS is at $k=\pi$. The spectral weights confirm that both bands have nearly equal $s$ and $p$ character (it is precisely equal for $\Delta=0$), except at $k=0$.  Indeed, at  $k=0$ the lower eigenstate is  of pure  $p$ character and is only visible in $A_{pp}$, while the upper eigenstate has pure $s$ character and is only visible in $A_{ss}$.

In the presence of EPC, the spectrum becomes much more complex. As typical in such problems, the lowest energy band is the narrow polaron band, above which there is a broadened continuum roughly following the dispersion of the bare bands, but also showing higher sidebands especially in the Peierls case.

Focussing on the polaron band, we see that for  Holstein coupling (bottom row) it  mimicks the character of the lowest bare band,  having nearly equal $s$ and $p$ character at the $X$ point, and only $p$ character at the $\Gamma$ point.  
On the other hand, a Peierls coupling (middle row) with the same strength produces a polaron state that has only $s$ character at the $\Gamma$ point, opposite to the bare lower band and the Holstein polaron band. The continuum above it also has much more structure than for the Holstein EPC, suggestive of multiple phonon side-bands.

Clearly, the Peierls coupling changes significantly the character of the lowest band. However, because there is already significant mixing of  $s$ and $p$ character for $g=0$, apart from the $k=0$ point it is not easy to understand how the Peierls coupling affects the character of the lowest eigenstate. This motivated us to consider the case  $t=0$: now there is  no mixing  between $s$ and $p$ orbitals when $g=0$, so any mixed character at finite $g$ originates from the Peierls EPC, giving a measure of its effects.

\begin{figure}[t]
	\centering
	\includegraphics[width=1\columnwidth]{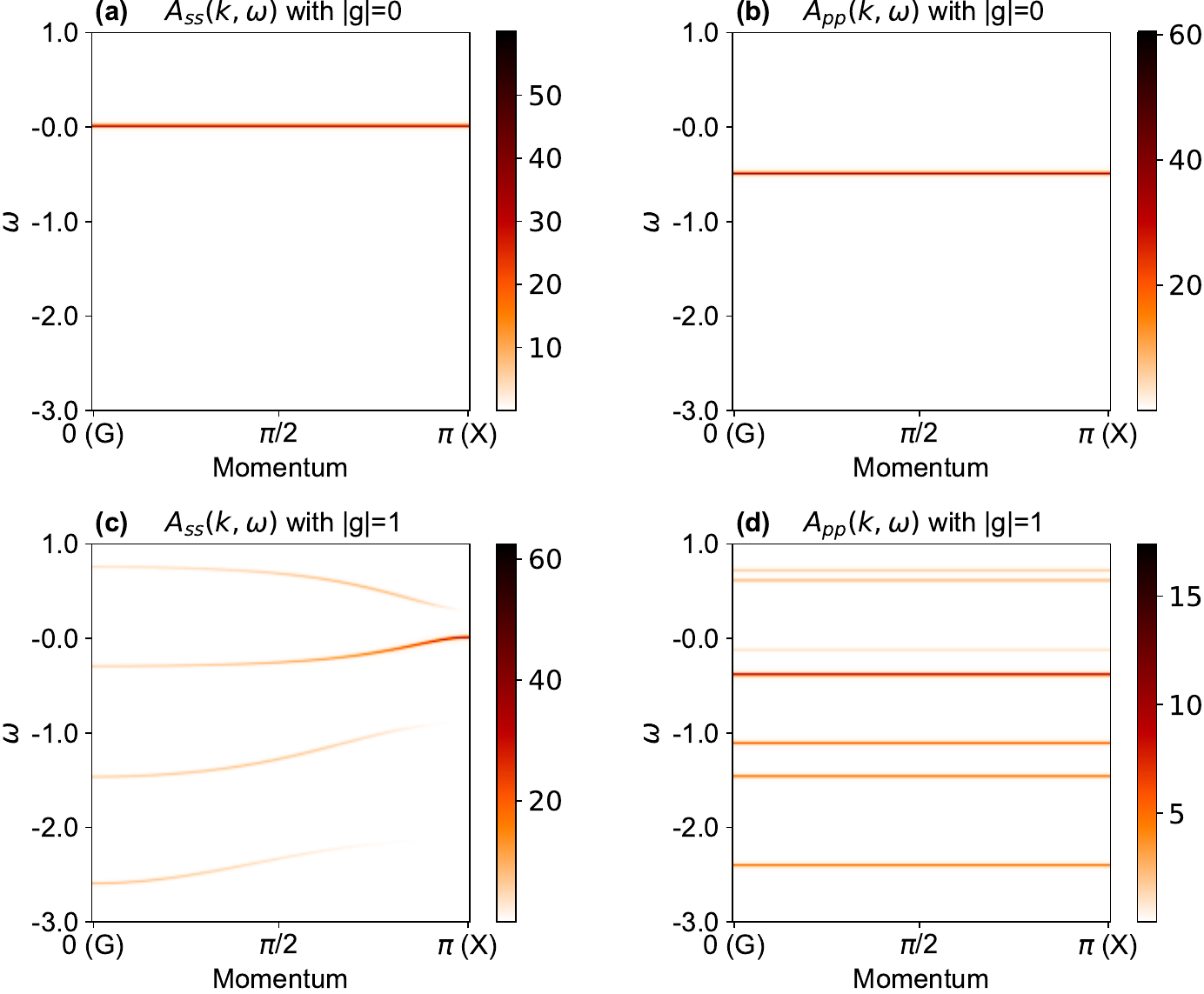}
	\caption{\label{sw_t_0} Spectral weights for the lowest states for a chain with  $t=0$ and (a-b) no electron phonon coupling; and (c-d) Peierls coupling. Other parameters  are the same as those in Fig. \ref{sw_t_not0}.}
\end{figure}

The $t=0$ spectral weights are shown in  Fig. \ref{sw_t_0}. The left column shows the spectral weights for $g=0$. As expected, there are two flat bands, one of pure $p$ character at energy $-\Delta$ and one of pure $s$ character at energy 0. The Holstein results look similar, except that the $p$-band moves to $-\Delta - g^2/\Omega$ (not shown). 

The right column in Fig. \ref{sw_t_0} shows the Peierls spectra with the same coupling strength as in Fig. \ref{sw_t_not0}. The first striking fact is that the two panels now show features at completely different energies, instead of having features at the same energies but with different weight as was the case for $t\ne 0$. This can be readily understood. Because the Peierls coupling switches the orbital character when a phonon is emitted or absorbed, a state with pure $s$-character only mixes with states that have $p$-character + one phonon (more generally, $p$+odd number of phonons and $s$+even number of phonons), {\em i.e.} these eigenstates are of the generic type $|P;k, s\rangle = (\phi_k s^\dagger_k + \sum_{q}^{} \phi_{k,q} p^\dagger_{k-q}b^\dagger_q +  \sum_{q, q'}^{} \phi_{k,q,q'} s^\dagger_{k-q-q'}b^\dagger_q b^\dagger_{q'} + \dots]|0\rangle$. Thus, there is zero probability of pure $p$-character, $\langle P; k, s| p_k^\dagger|0\rangle =0$, explaining why the eigenstates visible in $A_{ss}(k,\omega)$ are not visible in $A_{pp}(k,\omega)$, and viceversa.

  The second striking feature is that the eigenstates in the $s$-sector are dispersive, whereas those in the $p$-sector are still dispersionless. Regarding the former, and given that the bare bands are  dispersionless for $t=0$, the only possible way for momentum-dependence to arise is from a momentum-dependent $\Sigma(k,\omega)$. This confirms, again, that the Peierls self-energy is not local. Regarding the latter: the flat nature of the $p$ bands is because the Peierls coupling vertex depends only on the momentum of the $s$ operator, see Eq. (\ref{H_Pei}). A direct consequence of this is that all the sum rules $M^p_n \equiv \int_{-\infty}^{\infty} d\omega \omega^n A_{pp}(k,\omega)= \mel{0}{p_kH^np^{\dagger}_k}{0}$ are momentum independent when $t=0$. Indeed, it is straightforward to verify that  $M^p_0=1$, $M^p_1=-\Delta$, $M^p_2=\Delta^2+2g^2$, $M^p_3=-\Delta^3+2g^2\Omega-4g^2\Delta$, $M^p_4=\Delta^4+4g^2\Delta^2-2g^2\Omega\Delta+2g^2\Omega^2+12g^4$ and by induction to prove that momentum independence is observed to all orders $n$. This implies that $A_{pp}(k,\omega) \rightarrow A_{pp}(\omega)$ (when $t=0$) and therefore the spectrum in the $p$-sector must be dispersionless (when $t=0$ only). 

\begin{figure}[t]
	\centering
	\includegraphics[width=0.8\columnwidth]{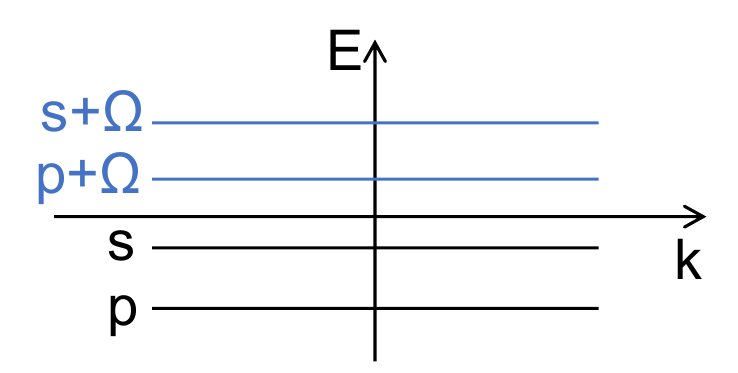}
	\caption{\label{curve_k_indp} Sketch of the low-energy spectrum when $t= g=0$: the states $p_k^\dagger|0\rangle$ have energy $-\Delta$, the states $s_k^\dagger|0\rangle$ have energy $0$, the states $p_{k-q}^\dagger b^\dagger_q|0\rangle$ have energy $-\Delta +\Omega$, the states $s_{k-q}^\dagger b^\dagger_q|0\rangle$ have energy $\Omega$, etc. }
\end{figure}

  The third striking feature of the Peierls spectra shown in  Fig. \ref{sw_t_0} is that the ground-state has $s$-character, even though in the absence of coupling, the $s$ states are energetically more expensive than the $p$ states and there is no bare hopping mixing them. The reason for this can be understood from Fig. \ref{curve_k_indp}, which sketches the low-energy spectrum when $t= g=0$. As already mentioned, the Peierls coupling hybridizes the lowest band of pure $p$ states with the continuum of $s$+one phonon states (and other, higher energy states). Level repulsion will push the lowest eigenstate in this sector  below the energy $-\Delta$ of the pure $p$ state; this energy lowering depends both on the strength of the hybridization (controlled by $g$) and on the split $\Delta+\Omega$ between the two manifolds. 

  Similar considerations apply in the $s$ sector, where the pure $s$ states mix with the $p$+one phonon states (and other, higher energy states). Their hybridization is again controlled by $g$, but the split between these states is now $|\Omega-\Delta|$. Because of this smaller split, the lowest eigenstate in the $s$ sector shifts down more than the one in the $p$-sector. Depending on the specific parameter values, this larger downward shift can place this eigenstate below the one in the $p$ sector, explaining why the ground-state in Fig. \ref{sw_t_0} is in the $s$-sector.

This argument can be made analytically more precise in limits where perturbation theory is valid. For instance, in the anti-adiabatic limit $\Omega \gg g, \Delta $, we can project out the high-energy states with one or more phonons, to find the effective Hamiltonian in the lowest energy (zero-phonon) manifold to be:
\begin{align*}
	\hat{h}=
	\sum_k \!
\left(
\begin{array}[c]{cc}
s^\dagger_k  & p^\dagger_k  \\
\end{array}
\right)\!\!
	\begin{pmatrix}
	-\frac{2g^2}{\Omega-\Delta}(1+\cos k) & 0 \\
	0 & -\Delta-\frac{2g^2}{\Omega+\Delta}
	\end{pmatrix}\!\!
        \left(\begin{array}[c]{c}
s_k\\
p_k\\
\end{array}
\right)
\end{align*}
This confirms that the lowest eigenstate in the $s$ sector is dispersive, while the one in the $p$-sector is flat. An intuitive understanding comes from the fact that through Peierls coupling, a carrier starting at an $s$ site can be moved to a neigbor $p$ site with a phonon created there, and then another Peierls process will absorb the phonon and can either bring the carrier to the original site or move it to the next $s$ site, thus giving rise to an effective $s$-$s$ nearest-neighbor hopping. By contrast, the lowest $p$-sector eigestate is dispersionless because a carrier starting at a $p$ site can  move to a neighbor $s$ site if a phonon is left behind at the original $p$ site. To return to the low-energy zero-phonon manifold, the next action of the Peierls coupling can only return the carrier back to the original $p$ site. Clearly, for the right combination of parameters, it is possible that $-\frac{4g^2}{\Omega-\Delta} < -\Delta-\frac{2g^2}{\Omega+\Delta}$, {\em i.e.} the finite-$g$ ground-state can move to $k=0$ in the $s$-sector, despite it being in the $p$-sector for $g=0$.

The effect of a finite $t$, within this antiadiabatic limit, is to add off-diagonal matrix elements of magnitude $2t \sin{(k/2)}$, leading to dispersive eigestates with mixed orbital character everywhere except at $k=0$. This analysis is in good agreement with the results shown for finite $t$ and supports our understanding of the overall behavior of the low-energy spectrum.

\subsection{The self-energy matrix}

\begin{figure}[t]
	\centering
	\includegraphics[width=1\columnwidth]{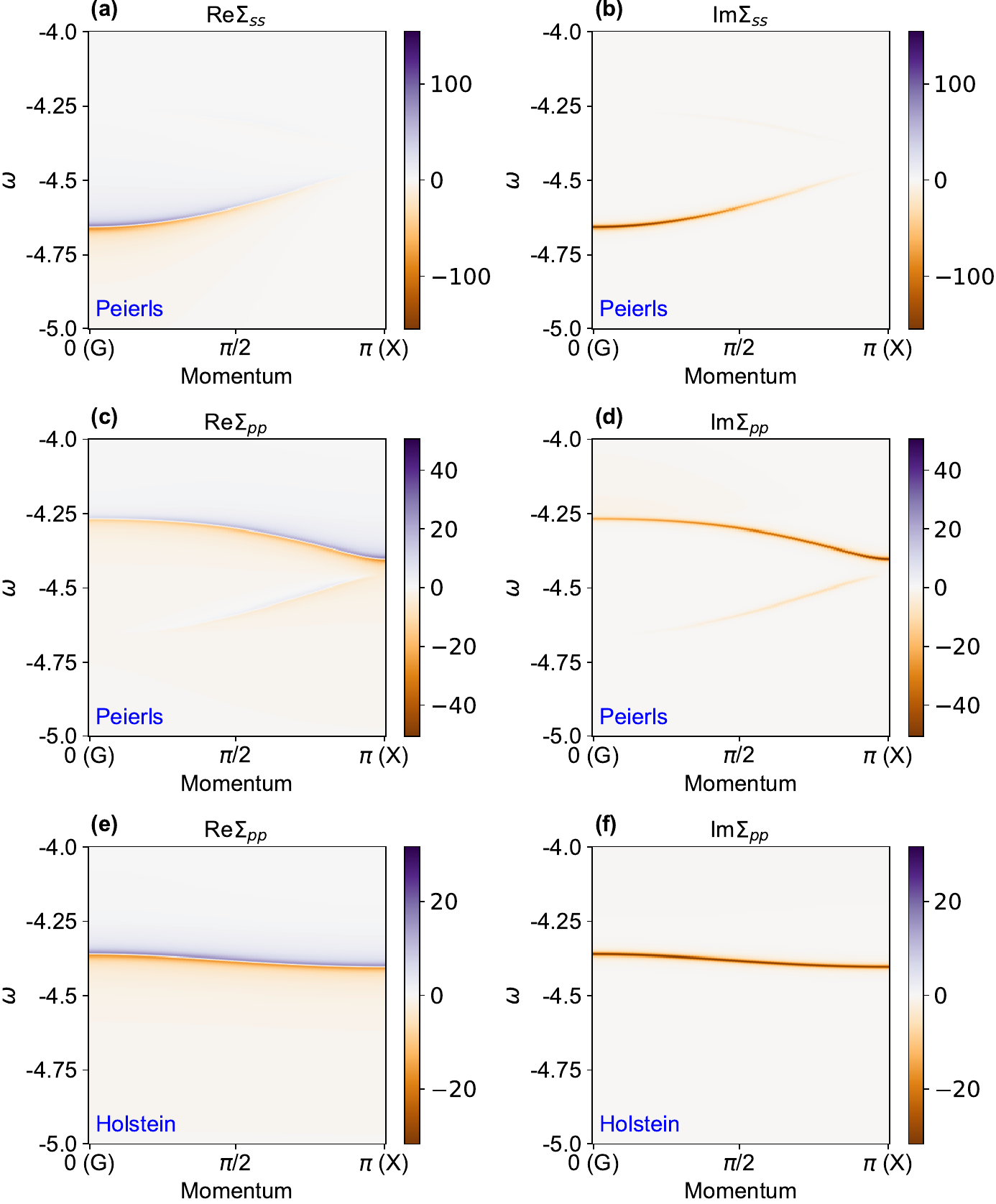}
	\caption{\label{SE_diag} (a)-(d) Contour plots of the real and imaginary parts of diagonal elements of the Peierls self energy as a function of momentum and energy $\omega$. The off-diagonal components are shown in the next figure. (e)-(f)  Contour plots of the real and imaginary parts of the Holstein $\Sigma_{pp}(k,\omega)$. We note that the Holstein $\Sigma_{ss}(k,\omega)\equiv 0$. Parameters are $\Omega=1$, $\Delta=0.5$, $t=2.5$, $g=1$ and $\eta= 5\times 10^{-5}$. }
\end{figure}

\begin{figure}[h]
	\centering
	\includegraphics[width=1\columnwidth]{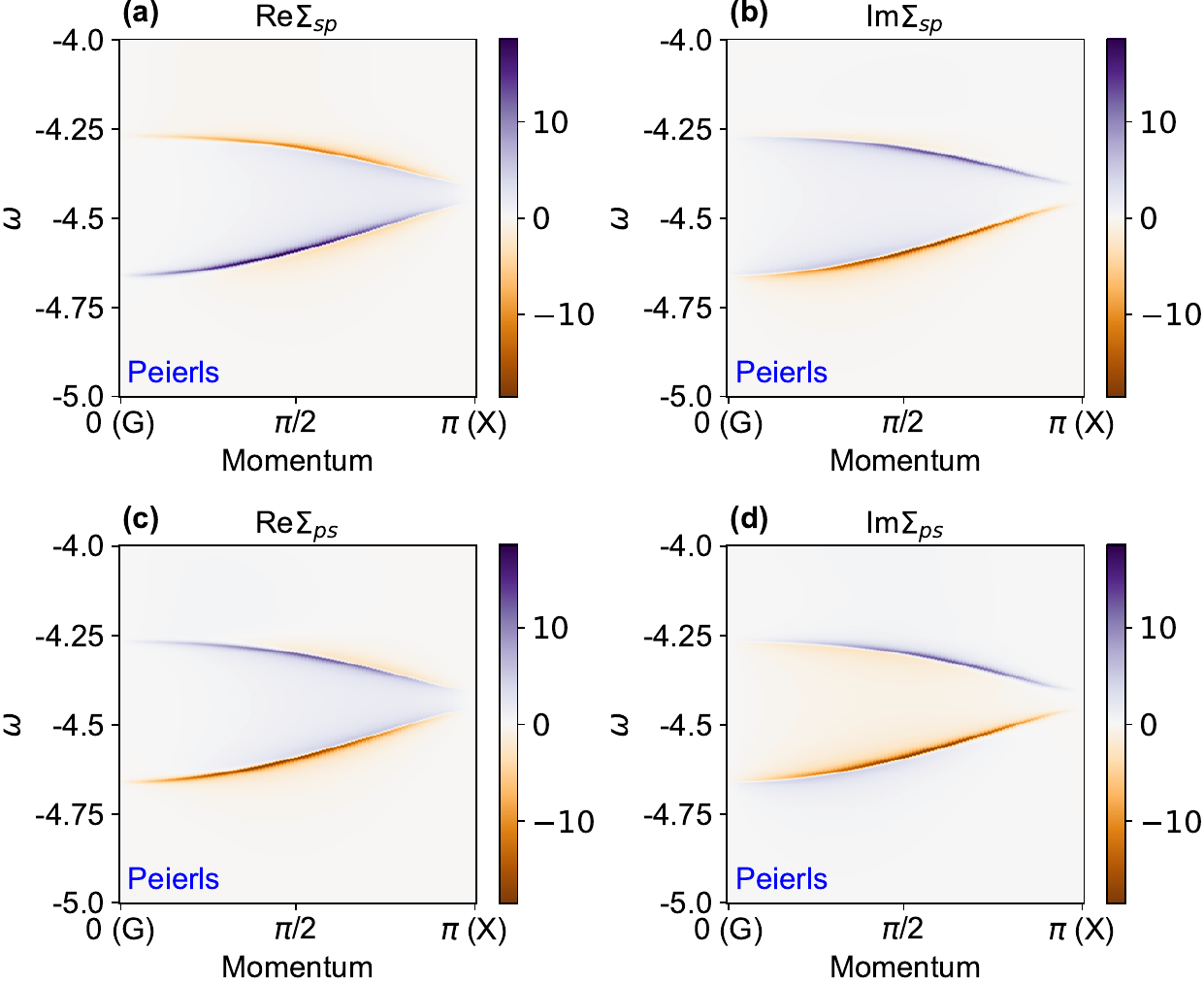}
	\caption{\label{SE_offdiag} Contour plots of the real and imaginary parts of the off-diagonal elements of the Peierls self energy  as a function of momentum and energy $\omega$. For the Holstein model, all off-diagonal elements vanish. Parameters are like in  Fig. \ref{SE_diag}.}
\end{figure}

We now plot the real and imaginary parts of the diagonal (in Fig. \ref{SE_diag}) and off-diagonal (in Fig. \ref{SE_offdiag}) elements of the self-energy vs. momentum and energy.  We show only their lowest energy feature(s), which determine the renormalization of the  lowest eigenstate (the polaron band). The panels are marked as `Peierls' or `Holstein' according to the corresponding model. Note that for the Holstein model, within MA we find $\Sigma_{ss}(k,\omega)= \Sigma_{sp}(k,\omega)= \Sigma_{ps}(k,\omega)=0$, hence they are not displayed. The parameters  $\Omega=1$, $\Delta=0.5$, $t=2.5$ and $g=1$ correspond to rather strong EPC, so as to make polaronic effects encoded in $\Sigma(k,\omega)$ more obvious; however, qualitatively similar results are obtained for other couplings.

For the Holstein model, Figs. \ref{SE_diag}(e),(f) show  the  real and imaginary parts of $\Sigma_{pp}(k,\omega)$ at low energies. We see that $\Sigma_{pp}(k,\omega)$ vanishes except in an extremely narrow range $\omega\in [-4.4,-4.3]$ where $\mbox{Re}\Sigma_{pp}$ changes sign while $\mbox{Im}\Sigma_{pp}$ has a negative peak. This narrow energy range is well above the unrenormalized GS energy $\epsilon_{GS}\approx -5.256$, and basically sets the location of the lower edge of the polaron+one-phonon continuum. Because this energy range with non-vanishing self-energy is so narrow, we conclude that the Holstein self-energy is essentially momentum independent within MA, {\em i.e.} $\Sigma_{pp}(k,\omega) \approx \Sigma_{pp}(\omega)$. Indeed, had we plotted the results against a broader energy range of order $10t$, comparable to the total bare bands widths, see Fig. \ref{1D_chain}(b), a momentum dependence would be undetectable. We conclude that with good accuracy, the Holstein model has a diagonal and local self-energy matrix.

The results for the Peierls self-energy are very different. Consider first the diagonal elements, shown in Figs. \ref{SE_diag}(a)-(d) at low energies. Here there are two visible features, a lower one starting at around $\omega=-6.65$ at $k=0$ and dispersing upwards, which is more visible in $\Sigma_{ss}$, and a higher one starting at around $\omega=-4.25$ at $k=0$ and dispersing downwards, more visible in $\Sigma_{pp}$. While the energy of these features varies more considerably with $k$ than in the Holstein model, this variation is still on a scale much smaller than the bare bandwith. However, unlike for the Holstein model, the {\em intensity} of this feature also has a very strong and non-trivial $k$ dependence, with the magnitude of $\Sigma_{ss}/\Sigma_{pp}$ decreasing/increasing significantly as we move from the center of the Brillouin zone ($k=0$, the $\Gamma$ point) to its edge ($k=\pi$, the $X$ point). This dependence is not an artifact of how the data is presented, but a real feature demonstrating non-trivial momentum dependence of the diagonal matrix elements. The off-diagonal Peierls self-energy matrix elements shown in Fig. \ref{SE_offdiag} also display both features, and also exhibit a very strong $k$-dependence of their intensities, with small intensities at $\Gamma$ and $X$, and a maximum around $k=\pi/2$.

These results show that for the Peierls coupling, {\em the self-energy matrix is not  local nor diagonal}. This is one of the main results of this work.

This is in stark contrast to the common belief in many spectroscopy experiments that one can treat the selfenergy as a scalar that is independent of momentum, and use that approximation to make predictions. Additionally, we  note that density functional calculations -- where the effect of the phonon is treated effectively as a static pseudopotential experienced by the one-particle electron wave function -- cannot tackle a dynamical coupling having different momenta, nor the dynamics in the selfenergy, not to mention the absence of the dressed particle. This could affect other properties like the EPC strength and corresponding transition temperatures between various phases, determined from such methods. 

Our results show that we cannot just assume the EPC and selfenergy to be independent of momentum. Furthermore, the  EPC involving different orbitals cannot be simply mapped into an effective one-band Holstein-like coupling, either. 

These conclusions align with previous results. For example, in perovskites like BaBiO$_3$, it is common to ignore the oxygen orbitals and focus only on the Bi valence orbitals as  generating the effective band \cite{Var1988, Has2007}. As pointed out by Khazraie et al.,\cite{Kha2018} however,  one has to include the oxygen orbitals in the model in order to explain its properties. We further investigated this in Ref. \onlinecite{Yam2020} and confirmed the qualitatively different properties of the Peierls and effective Holstein models in perovskite-like lattices. As another example, the one-band $t$-$J$ model is a very popular way to model cuprates. Ebrahimnejad et al. \cite{Ebr2014}, however, showed that the less simplified three-band model, which includes the oxygen orbitals explicitly, predicts qualitatively different physics. The same remains true if coupling to phonons is also included beyond the simplistic Holstein model.\cite{Yam2020} All of these examples point to the importance of explicitly including all the various relevant orbitals when describing EPC.

\subsection{\label{Spec}Angle resolved photoemission intensity}

A natural question is whether these differences between the self-energies associated with the two different types of EOC could be inferred  experimentally. The experimental technique that is most directly used to infer self-energies is Angle Resolved Photoemission Spectroscopy (ARPES). To investigate this question, we consider a $s$-$p$ chain with smaller (and more physical) coupling of $g=0.5$.

Instead of projecting again on the $s^\dagger_k|0\rangle, p^\dagger_k|0\rangle$ basis, now we project on the  eigenstates of the bare Hamiltonian ${\cal H}_0|l, k\rangle = \epsilon_{low}(k) |l,k\rangle$ and  ${\cal H}_0|t, k\rangle = \epsilon_{top}(k) |t,k\rangle$, where $l$ and $t$ label the low and top bare bands, respectively, $\epsilon_{low}(k) < \epsilon_{top}(k)$ (see top row of Fig. \ref{sw_t_not0}). This projection produces another representation of the $2\times2$ Green's function matrix, {\em e.g.}  $ G_{ll}(k,\omega) = \langle l, k| \hat{G}(\omega) |l, k\rangle$  and its spectral weight $A_{low}(k, \omega) = - {1\over \pi} Im G_{ll}(k,\omega)$, {\em etc}.

\begin{figure}
	\centering
	\includegraphics[width=1\columnwidth]{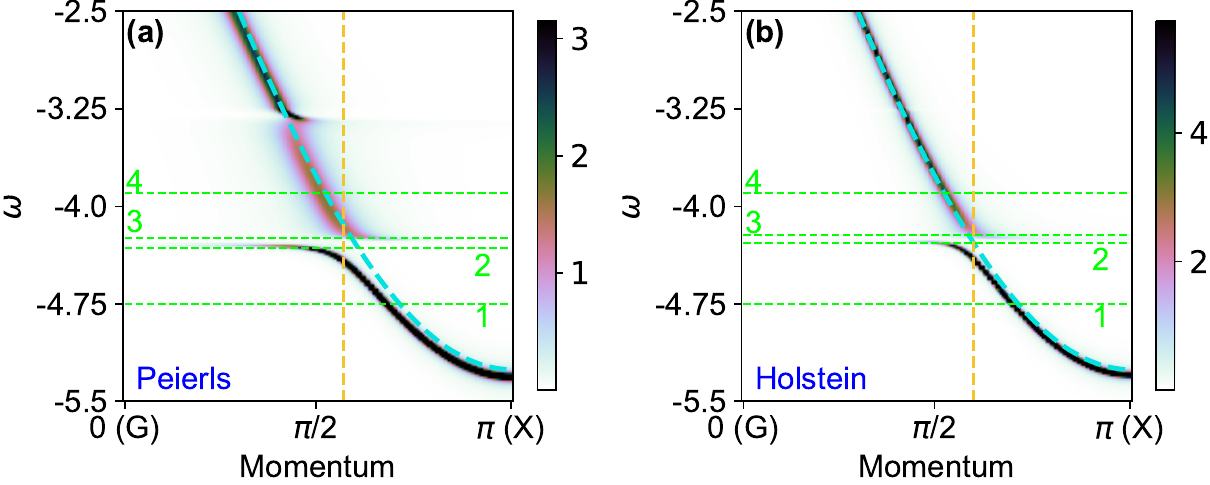}
	\caption{\label{Pei_Hol_overlay} Low-energy range of the spectral weight $\tanh A_{low}(k,\omega)$ for (a) Peierls, and (b) Holstein EPC. All parameters are as before, except for a smaller $g=0.5$. 
         Pale blue dashed curves indicate the dispersion of the lowest bare band. The vertical orange  dotted lines are the eye guides along a constant momentum cut,  and the four green horizontal dotted lines labelled 1-4 are eye guides for  different constant energy cuts.}
\end{figure}

Figure \ref{Pei_Hol_overlay} shows the spectral weight $A_{low}(k, \omega)$ in the range of the lowest-energy  states, for Peierls (left) and Holstein (right) EPC. As expected, in both cases the EPC leads to a continuum appearing at energy $\Omega$ above the bottom of the polaron band; the latter flattens just below this continuum as $k\rightarrow 0$. The weight in the continuum is roughly centered around the bare dispersion (shown by dashed line), and shows more broadening in the Peierls than in the Holstein case. This broadening reflects the fact that these states have high enough energy to emit a real phonon, hence their finite lifetime.

In ARPES analysis, it is customary to extract information regarding the self-energy either from cuts at constant momentum (like the vertical orange lines), or cuts at constant energy (like the green horizontal lines). Constant momentum cuts show a single peak below the continuum, confirming the existence of the polaron state. For the same value of $g$, the polaron band lies at lower energies in the Peierls than in the Holstein case. Combined with the broader continuum and the feature at $2\Omega$, this shows that the effective EPC is larger in the Peierls than in the Holstein case, for the same value of $g$.

The four horizontal green lines labelled 1-4 in Fig. \ref{Pei_Hol_overlay} (a),(b) illustrate  constant energy cuts. For energies well below the continuum (line 1), this cut has a single narrow peak, marking the infinitely-lived polaron.  Just below the continuum (line 2), the peak becomes much broader, due to the flattening of the polaron band. As it enters the continuum (line 3), the broadening expands further but now towards the $X$ point. Moving to higher energy, the peak  narrows somewhat again, but the associated lifetime is here always finite. 

Typical ARPES analysis infers the imaginary part of the self energy $\Im \Sigma_{\tilde{\omega}}$ from the measurement of spectral weight intensity, assuming the self energy $\Sigma_{\tilde{\omega}}$ is independent of momentum. With this assumption, the spectral weight $A_{\tilde{\omega}}(k)$ on the constant energy $\tilde{\omega}$ cut becomes  a Lorentzian as a function of the momentum $k$: \cite{Nei2007,Vee10,Vee11}
\begin{align*}
	A_{\tilde{\omega}}(k)=-\frac{1}{\pi}
	\frac{\frac{\Im\Sigma_{\tilde{\omega}}}{\epsilon'(k_m)}}
	{(k-k_m)^2+(\frac{\Im \Sigma_{\tilde{\omega}}}{\epsilon'(k_m)})^2}
\end{align*}
where $\epsilon(k)$ is the dispersion of the  bare band, $\epsilon'(k)$ is its derivative with respect to $k$ and the $k_m$ is the momentum where the spectral weight peaks on the cut with  energy $\tilde{\omega}$. Reading from the Lorentzian-form expression of $A_{\tilde{\omega}}$, we can infer  as a result:
\begin{align*}
\Im \Sigma_{\tilde{\omega}}=- \text{HWHM}\cdot\epsilon'(k_m)
\numberthis\label{Im_Sigma}
\end{align*}
where $\text{HWHM}$ is the Half Width at Half Maximum of the peak.

\begin{figure}[t]
	\centering
	\includegraphics[width=1\columnwidth]{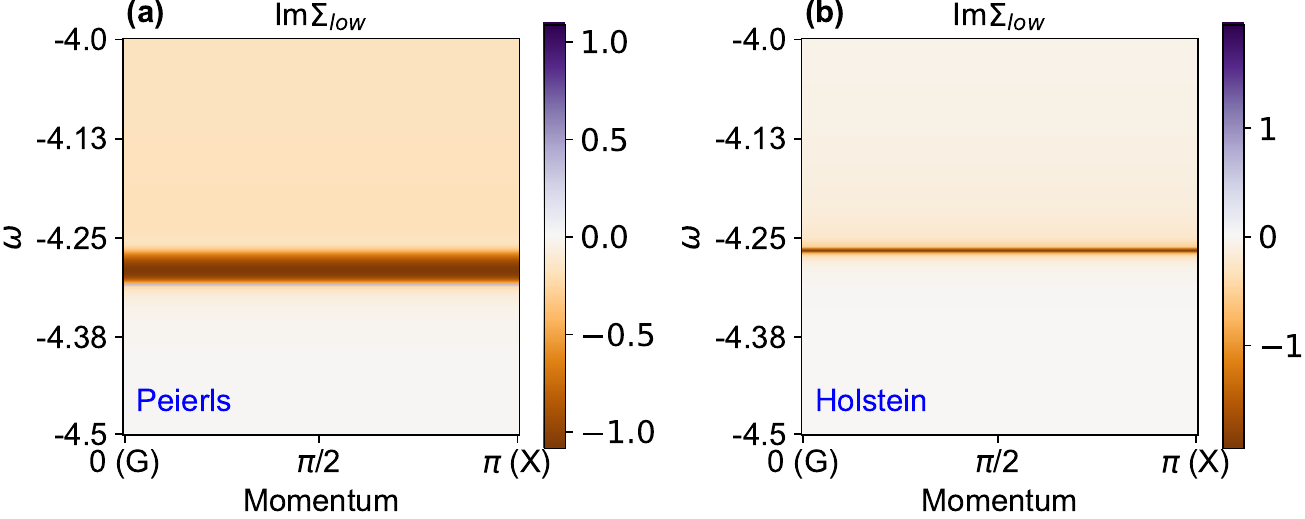}
	\caption{\label{ARPES_SE} Contour plots of the imaginary part of the 'ARPES selfenergies' extracted from Eq. (\ref{Im_Sigma}) and the constant energy cuts of the spectral weights shown in Fig. \ref{Pei_Hol_overlay} for  (a) Peierls, and (b) Holstein EPC.}
\end{figure}

\begin{figure}[t]
	\centering
	\includegraphics[width=1\columnwidth]{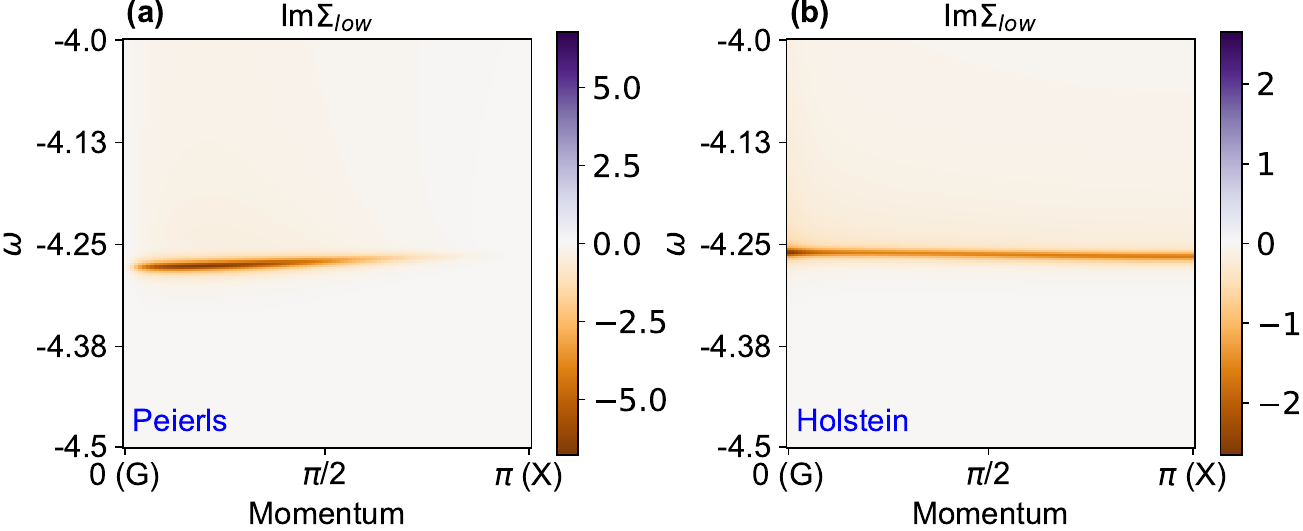}
	\caption{\label{MA_SE} Contour plots of the imaginary parts of the  selfenergy ($\Im\Sigma_{\text{low}}$) that generated the spectral weights in Fig. \ref{Pei_Hol_overlay} for (a) Peierls, and (b) Holstein EPC.}
\end{figure}

We use this analysis on the constant energy cuts obtained from the spectral weights shown in Fig. \ref{Pei_Hol_overlay}, to infer their corresponding  'ARPES self-energies'; these are shown in Fig. \ref{ARPES_SE}. For comparison, Fig. \ref{MA_SE} shows the imaginary part of the actual self-energies $\Sigma_{\text{low}}$ that generated the spectral weights in Fig. \ref{Pei_Hol_overlay}.

For the Holstein coupling (right panels), the 'ARPES selfenergy' agrees well with the actual one: both have a dispersionless peak at $\omega\approx-4.25$. This confirms that this ARPES analysis works well for local self-energies. On the other hand, as already discussed, the self-energy in the Peierls model is momentum dependent (left panel of Fig. \ref{MA_SE}). The ARPES analysis, however, by construction still extracts a local self-energy (left panel of Fig. \ref{ARPES_SE}). Not only is the momentum-dependence of the selfenergy totally missed, but there is no clear sign that the ARPES analysis is failing: the peaks in the constant energy cuts are well approximated by Lorentzians in the Peierls case as well. This exemplifies that what appears to be a successful analysis with the typical ARPES approach can produce very wrong results.

In the discussion above, we used the diagonal element $\Sigma_{\text{low}}(k,\omega)$ projected on the lower bare eigenstate $|l, k\rangle$ for the comparison with the 'ARPES selfenergy'. As already noted, the true self-energy is a $2\times2$ matrix and it is not apriori clear which combination of its matrix elements should be used for this comparison. However, for Peierls EPC,  all of them have non-trivial momentum dependence and therefore so would  the correct combination.  Our particular choice suffices for the purpose of proving the lack of reliability of the usual ARPES analysis when the self-energy is non-local.

We now proceed to find the proper link between the $2\times2$ Green's function matrix and the ARPES intensity, to illustrate further challenges in the analysis of ARPES data. As already mentioned in the Introduction, for  systems with a single band, the ARPES intensity  $I_{exp}({\bf K},\omega) \propto |M({\bf K,k})|^2 A({\bf k},\omega) f(\omega)$ is directly proportional to the spectral weight associated with that band, up to a matrix element and a Fermi-Dirac function. For a full band like we consider,  $f(\omega)\equiv 1$ and we  ignore it.

For a multi-band system like the $s$-$p$ chain we are investigating, we need to generalize this formula accordingly.  We follow the usual steps sketched in Refs. \onlinecite{Dam2003,Day2019,Day2020}. The system is initially in its $N$-electrons ground-state $|GS\rangle$. Upon absorbing a photon of energy $\omega$ and momentum ${\bf q}$, it emits a photo-electron of momentum ${\bf K}$ and energy  $\epsilon_K=\hbar^2K^2/2m$ and transitions into an eigenstate $|\Psi^{(\alpha)}_{N-1}\rangle$ of the $N-1$ electron system, {\em i.e.} the possible final states are $|{\bf K},\alpha\rangle=|f,{\bf K}\rangle\otimes|\Psi^{(\alpha)}_{N-1}\rangle$, where the first ket describes the photo-electron $\langle {\bf r}|f, {\bf K}\rangle\propto \exp(i {\bf K}{\bf r})$. According to  Fermi's Golden Rule, the corresponding probability is:
\begin{align*}
w_{{\bf K},\alpha} = 2\pi |\langle {\bf K},\alpha| {\cal H}_{\rm int}|GS\rangle|^2 \delta(\epsilon_{{K}} +E^{(\alpha)}_{N-1} - E^{GS}_N-\omega)
\numberthis
\end{align*}


Using the completeness of the one-particle basis, we can rewrite $|GS\rangle =\sum_{k}^{} [|s_{k}\rangle \otimes s^\dag_{-k}|GS\rangle + |p_{k}\rangle \otimes p^\dagger_{-k}|GS\rangle]$. We remind the reader that for us, $s^\dagger_{-k}$ creates a hole (ie, removes an electron) with momentum $-k$ from the chain; here $|s_k\rangle$ is the Bloch state for an electron with momentum $k$.  We can then re-write:
\begin{align*}
  \langle {\bf K},\alpha| {\cal H}_{\rm int}|GS\rangle& = \sum_{k}^{}\big[\langle f,{\bf K}|{\cal H}_{\rm int}|s_{k}\rangle \langle \Psi^{(\alpha)}_{N-1}| s^\dagger_{-k}|GS\rangle \\
  &+ \langle f,{\bf K}|{\cal H}_{\rm int}|p_{k}\rangle \langle \Psi^{(\alpha)}_{N-1}| p^\dagger_{-k}|GS\rangle\big] \\
\numberthis
\end{align*}
Note that for a single band we would have a single operator  $c_k$ with momentum along the chain $k$, thus only one term would appear on the rhs, in agreement with Ref. \onlinecite{Dam2003}. Doing all the standard manipulations, we arrive at our final result:
\begin{align*}
 &I_{\rm exp}({\bf K},\omega)\propto	-{1\over \pi} \mbox{Im}\big[\sum_{k} 
\left(
\begin{array}[c]{cc}
 M^*_s({\bf K},k)& M^*_p({\bf K},k)   \\
\end{array}
\right)\\&\times
	\begin{pmatrix}
	G_{ss}(k,\omega) & G_{sp}(k,\omega) \\
 	G_{ps}(k,\omega) & G_{pp}(k,\omega)
	\end{pmatrix}\left(
\begin{array}[c]{c}
M_s({\bf K},k)\\
M_p({\bf K},k)\\
\end{array}
\right)\big]
 \numberthis\label{I_exp}
\end{align*}
The dipole matrix elements  $M_s({\bf K},k) = \langle f, {\bf K}| {\cal H}_{\rm int}|s_{k}\rangle$, $M_p({\bf K},{k}) = \langle f, {\bf K}| {\cal H}_{\rm int}|p_{k}\rangle$  are associated with the two possible Bloch states formed from $s$ and $p$ orbitals, respectively. We note that we could have used instead the basis diagonalizing ${\cal H}_0$, which leads to replacing indexes $s,p \rightarrow l, t$ in the equation above, the corresponding matrix elements being $M_{l/t}({\bf K},k) = \langle f, {\bf K}| {\cal H}_{\rm int}|l/t, k\rangle$. Because they are related through a unitary transformation, both expressions give the same total $I_{\rm exp}({\bf K},\omega)$. 

\begin{figure*}
	\centering
	\includegraphics[width=\textwidth]{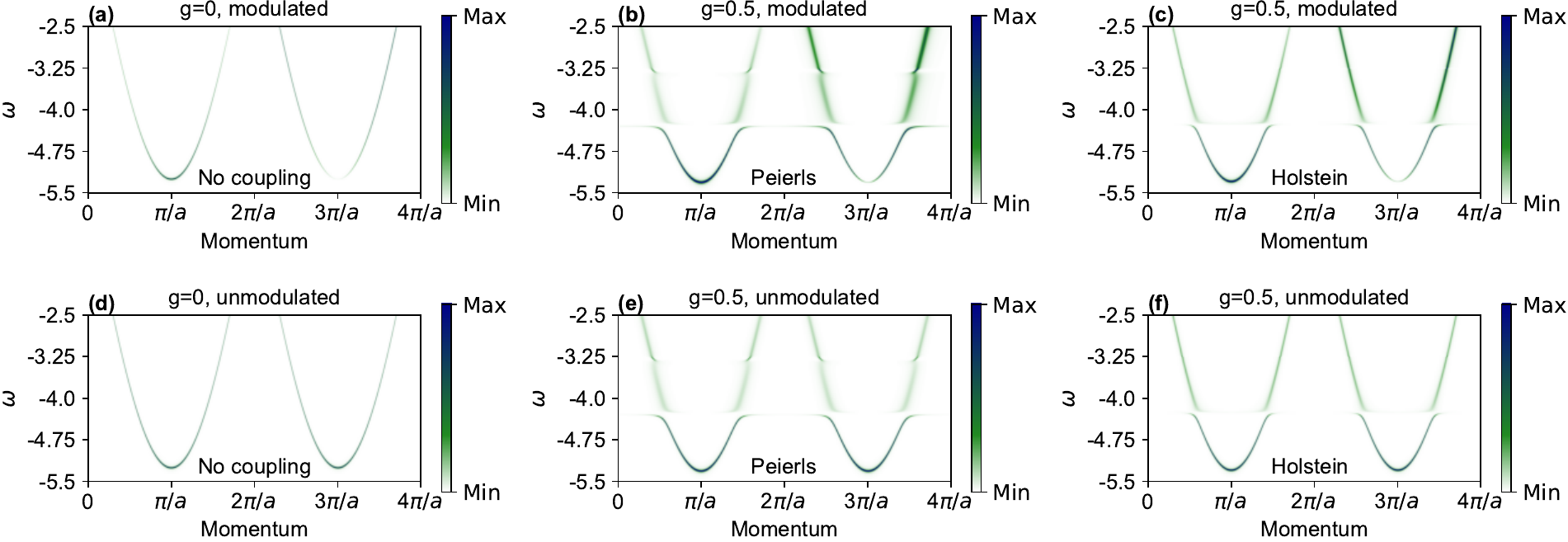}
	\caption{\label{I_exp_plot} Experimentally relevant $\tilde{I}_\text{exp}$ calculated from Eq. (\ref{mat})  for (a) no coupling, (b) Peierls coupling, and (c) Holstein coupling. Here we set the lattice constant $a=4.5\AA$, and the Bohr radia are chosen to be $1.17\AA$ and $0.63\AA$ for the $s$ and $p$ orbitals, respectively. The photon energy  is taken to be 70eV with a  workfunction $W=4.22$eV;  all other parameters are as in Fig. 4. The top three panels show the result of Eq. (\ref{tI_exp}) with the matrix elements calculated as described in the text, whereas the bottom panels show the results when we set all matrix elements to unity. }
\end{figure*}

Ignoring the momentum $|{\bf q}|\ll |{\bf K}|$ of the photon, the discrete translational invariance of ${\cal H}_{\rm int}$ along the chain direction -- taken to be the $x$ axis -- requires that $k= K_x$ (mod $2\pi$), ie the photo-electron has the same momentum projection along the chain as the original electron that absorbed the photon, up to a reciprocal lattice vector. This removes the sum over $k$ from Eq. (\ref{I_exp}). To make further progress, we will employ some rather drastic approximations, which nevertheless are widely used in the literature. We comment on the implementation of more accurate approaches below. Within the dipole approximation ${\cal H}_{\rm int} = {\bf A}({\bf r})\cdot {\bf p} \approx \bm{\epsilon}\cdot {\bf p}$ where $\bm{ \epsilon}$ is the light polarization, leading to:
  \begin{align*}
    M_{s/p}({\bf K},k)\propto \sum_{n}^{}\delta_{K_x,k+2\pi n}  (\bm{\epsilon}\cdot {\bf K}) 
    m_{s/p}({\bf K})
    \label{mat}
\numberthis
\end{align*}
  where
  \begin{align*}
   m_{s/p}({\bf K})=e^{-iK_x \delta_{s/p}}\int_{}^{}d \bm{\rho} e^{-i {\bf K}\cdot \bm{\rho}} \phi_{s/p}(\bm{\rho})
    \label{msp}
\numberthis
\end{align*}
  with $\delta_s=0, \delta_p=a/2$ marking the location of the orbitals inside the unit cell, and $\phi_{s/p}(\bm{\rho})$  being the $s/p$ Wannier orbital centered at the origin. It is important to note that the phase factors $e^{-iK_x \delta_{s/p}}$ appear because of our choice, in Eq. (\ref{ops}), to reference all the Bloch states to the same location $R_i$, instead of using $R_i$ for $s$-orbitals, and $R_i+a/2$ for $p$-orbitals. Had we made the latter choice, this phase-factor would disappear from Eq. (\ref{msp}) but would be precisely compensated for by additional phase-factors in the $G_{sp}(k,\omega), G_{ps}(k,\omega)$ offdiagonal matrix elements, and $I_{\rm exp}({\bf K},\omega)$ in Eq. (\ref{I_exp}) remains the same. For convenience, in the discussion below we will refer to the phase factor $e^{-iK_x (\delta_{s}-\delta_p)}= e^{iK_x a/2}$, which appears because the $s$ and $p$ orbitals are located at different positions in the unit cell, as the 'interference factor'.

  For simplicity,  we approximate the Wannier orbitals as  atomic-like  orbitals with the correct symmetry and reasonable values for the corresponding Bohr radii, which allows us to calculate the matrix elements analytically. While a more accurate description would likely be necessary for analysis of actual experimental data, this very simplified choice suffices to help us illustrate our points.

With these approximations, for $1s$ and  $2p_x$ orbitals we find:
\begin{align}
 m_s(\bm{K})=\frac{8\sqrt{\pi}}{a^{5/2}_B}\frac{a_B^4}{[1+(a_B K)^2]^2} \label{ms} \numberthis \\
   m_{p}({\bf K})= -i e^{-iK_x {a\over 2}} \sqrt{\pi Z a_B \over 2} {K_x(16a_B)^2Z^3\over [Z^2+ 4 (a_BK)^2]^3}
    \label{mp}
\numberthis
\end{align}
Note that the $a_B$ values entering the two formulae are those appropriate for the corresponding orbitals; we call both $a_B$ simply for convenience. We also note that in the results shown below, we used the $m_s(\bm{K})$ for a $6s$ orbital because the parameters we are using are appropriate for a BiO chain. This latter matrix element is a lot more complicated than the one for the $1s$ orbital listed above, so we do not write it here.

Before continuing, it is important to emphasize that a more careful calculation of the matrix elements is likely to generate even stronger, and even more different momentum-dependence for the two Bloch states. The calculation sketched above has two significant shortcomings: (a) it replaces Wannier orbitals with atomic orbitals, and (b) it replaces the high energy eigenstate describing the photo-electron by a simple plane wave. Regarding (a), clearly atomic orbitals placed at different sites are not orthogonal, so they cannot be a reasonable approximation for the Wannier function. Fixing this is possible  \cite{Der2022} but leads to much more complicated spatial profiles which will add non-monotonic momentum-dependent features to the matrix elements. Regarding (b), the high-energy eigenstates must be orthogonal to the low-energy ones, so the photoelectron state cannot be a plane wave. Replacing it with a more accurate Bloch state means that the prefactor in the matrix element is more complicated than $\bm{\epsilon}\cdot {\bf K}$, and the integrand itself will have additional ${\bf K}$ dependence, further complicating the momentum-dependence of the matrix elements. For more details on these issues, see Refs. \onlinecite{Day2020, Day2019}.

Nevertheless, continuing with our simplifying approximations, we can factor out $\bm{\epsilon}\cdot {\bf K} $ from both matrix elements to write $I_{\rm exp}(\bm{K},\omega)= (\bm{\epsilon}\cdot {\bf K})^2 \tilde{I}_{\rm exp}(\bm{K},\omega)$ where 
  \begin{align*}
 &\tilde{I}_{\rm exp}({\bf K},\omega)\propto -{1\over \pi} \mbox{Im}\big[
\left(
\begin{array}[c]{cc}
 m^*_s({\bf K})& m^*_p({\bf K})   \\
\end{array}
\right)\\&\times
	\begin{pmatrix}
	G_{ss}(K_x,\omega) & G_{sp}(K_x,\omega) \\
	G_{ps}(K_x,\omega) & G_{pp}(K_x,\omega)
	\end{pmatrix}\left(
\begin{array}[c]{c}
m_s({\bf K})\\
m_p({\bf K})\\
\end{array}
\right)\big]
 \numberthis\label{tI_exp}
  \end{align*}
  The overall prefactor $(\bm{\epsilon}\cdot {\bf K})^2 $ depends only on the geometry of the experiment, but its very simplified form is a direct consequence of the approximations mentioned above. In particular, the approximation of describing the photoelectron with a plane-wave does not properly take into account the $\Delta l = \pm 1$ selection rule required in the dipole approximation. This can be achieved by expanding the plane-wave in spherical harmonics and selecting only the components satisfying the selection rule, when calculating the dipole matrix element, as done in Chinook \cite{Day2019,Day2020}. Doing this would result in different prefactors and momentum dependences for the contributions from the various $s$ and $p$ components, in the equation above, making it even more complicated. We note, however, that if there are  high symmetry points where the wavefunctions have pure orbital character (as is the case for our model at the $\Gamma$ point, in the absence of electron-phonon coupling), then at those momenta the equation above simplifies signficantly because the off-diagonal propagators vanish. In such special cases, it may be possible to measure separately the diagonal contributions.  Generically, though, we expect contribution from all orbitals to be mixed together, with additional $k=K_x \phantom{a} (\mbox{mod} \phantom{a}2\pi/a)$ dependence coming from the dipole matrix elements, as suggested by Eq. (\ref{tI_exp}).

 This generic situation is illustrated in Fig. \ref{I_exp_plot}, where in the top panels we plot  $\tilde{I}_{\rm exp}({\bf K},\omega)$ of Eq. (\ref{tI_exp}) versus $K_x>0$ up to the third Brillouin zone, when we calculate the matrix elements from Eqs. (\ref{ms}), (\ref{mp}). Note that the value of  $K = \sqrt{K_x^2+K_\perp^2}$  is found from conservation of energy; the results shown in Fig. \ref{I_exp_plot}  assume an incoming photon with an energy of 70 eV and a workfunction of 4.2eV. The value of ${\bf K}_{\perp}$ is irrelevant for the Green's functions contributions,  but it affects the matrix elements.

 Looking first inside the first Brillouin zone $K_x\le \pi/a$, we find that the location of the main features remains the same as in  Fig. \ref{Pei_Hol_overlay}, because the spectra are the same. However, the relative spectral weights are changed by the weighted average in Eq. (\ref{tI_exp}). 
  We could now repeat the ARPES analysis for these spectral weights, and the results would be similar to those in Fig. \ref{ARPES_SE}: the built-in assumption of a local self-energy would  result again in the 'prediction' of a local selfenergy, even for the Peierls case where this is obviously not correct. 

 We can now illustrate clearly the challenge of extracting the selfenergy from the ARPES intensity for a multi-band system. Consider the already very simplified Eq. (\ref{tI_exp}), and let us make the additional -- totally unjustified -- assumption that the matrix elements of both orbitals are the same, $ m_s({\bf K})=m_p({\bf K})$, so that an overall prefactor $|m_s({\bf K})|^2$ can be pulled in front of the expression like in one-band systems, to find that $I_{\rm exp}({\bf K},\omega) \propto \mbox{Tr} [G(K_x, \omega)]$ depends on the {\em  trace} of the $2\times2$ matrix of Green's functions. We could use ab-initio methods to calculate the matrix of bare Green's functions $[G_0(K_x,\omega)]$ in the same orbital basis, however only having access from ARPES to the trace of $[G(K_x,\omega)]$ does not provide sufficient information to allow us to calculate the various elements of the matrix  $[\Sigma(K_x,\omega)]$ using Eq. (\ref{e5}). Needless to say, in the realistic case with  $ m_s({\bf K})\ne m_p({\bf K})$, the task of extracting an accurate selfenergy from the ARPES intensity becomes even more difficult, if not  hopeless.

\begin{figure*}
	\centering
	\includegraphics[width=\textwidth]{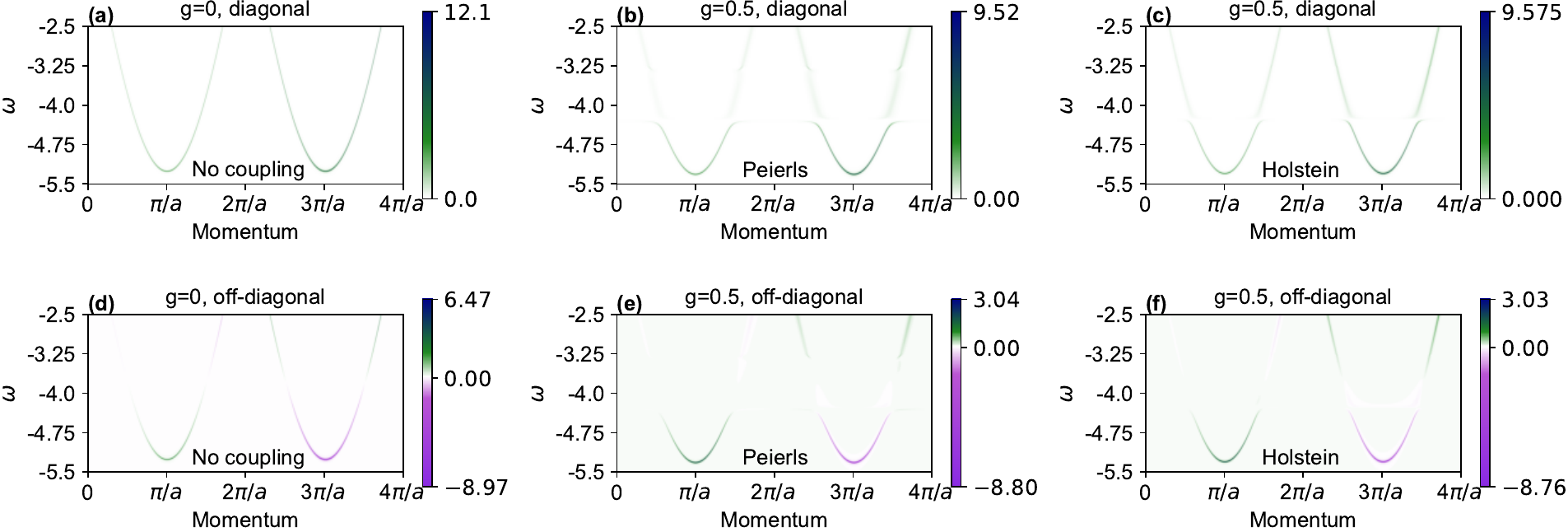}
	\caption{\label{I_exp_plot2}  Experimentally relevant $\tilde{I}_\text{exp}$ calculated from Eq. (\ref{mat})  for (a) no coupling, (b) Peierls coupling, and (c) Holstein coupling. Here we set the lattice constant $a=4.5\AA$, and the Bohr radia are chosen to be $1.17\AA$ and $0.63\AA$ for the $s$ and $p$ orbitals, respectively. The photon energy is taken to be 70eV and the workfunction is 4.22eV, all other parameters are as in Fig. 4. The top three panels show the result of Eq. (\ref{tI_exp}) with the matrix elements calculated as described in the text, whereas the bottom panels show the results when we set all matrix elements to unity. }
\end{figure*}

 Fig.  \ref{I_exp_plot} shows the estimated 'ARPES' weight also outside the first Brillouin zone $|K_x|>\pi/a$, in order to illustrate the  additional modulation of the spectral weights between different Brillouin zones due to the inclusion of the matrix elements. All three top panels show a vanishing weight  for the lowest energy feature at $K_x=3\pi/a$, even though this feature is clearly visible at $K_x=\pi/a$. This modulation is entirely due to the matrix elements, as evident from the fact that it appears for three different Hamiltonians. Graphically, this is demonstrated in the bottom panels of Fig.  \ref{I_exp_plot}, which correspond to $m_s({\bf K})= m_p({\bf K})=1$ resulting in  'ARPES' spectra that have the Brillouin zone periodicity because all  $G_{ss'}(K_x,\omega)$ propagators have this periodicity.

 There are two contributions to this modulation of $\tilde{I}_{exp}({\bf K},\omega)$. The first comes from the fact that the magnitude of the $m_{s/p}({\bf K})$ changes differently with ${\bf K}$, see Eqs. (\ref{ms}), (\ref{mp}), and therefore their ratio differs for values of $K_x=k + n {2\pi\over a}$ that map onto the same $k$ value.  The second is due to the 'interference factor' $\exp(i K_x a/2)$ mentioned before, which is a consequence of the different locations of the $s$ and $p$ orbitals within the unit cell.

 We can gauge their relative importance by comparing the contribution from the diagonal matrix elements vs. that from the off-diagonal matrix elements to Eq. (\ref{tI_exp}); only the latter is sensitive to the 'interference' factor, but both are sensitive to the monotonic change with ${\bf K}$ of the ratio of the matrix elements. This comparison  is  shown in  Fig. \ref{I_exp_plot2}, where the top plots show  the diagonal contributions, while the bottom show  the off-diagonal ones (the sum of the two sets gives the total results plotted in the top line of Fig. \ref{I_exp_plot}). The top plots show some variation of the results between the different Brillouin zones, for example the higher energy continuum is bigger for the larger $K_x$ values than for their counterparts $K_x-2\pi/a$. As expected, this change is monotonic. The much more spectacular change is seen in the contribution from the off-diagonal matrix elements, whose contribution changes very significantly between consecutive Brillouin zones as the 'interference factor'   $\exp(i K_x a/2)$ changes sign from $+i$ to $-i$ for $K_x=\pi/a $ vs. $3\pi/a$, resulting in $m_p({\bf K})$ changing from a positive to a negative real number at these particular momenta, see Eq. (\ref{mp}).

 \section{\label{sec:discuss}Discussion}

 In this work, we studied the simplest 1D two-band model with two types of EPC in an effort to understand the link between theoretically calculated Green's functions and self-energies -- which are matrices in multi-band systems -- and the ARPES intensity, which is a scalar. Our simple model avoids the additional complication typical in more complex multi-band system, of having 'spaghetti-like' bands overlapping in the same range of energies. Instead, our model has a lower and an upper band separated by a gap (in the absence of EPC). For the weak EPC studied in the second part of the work, one would normally assume that only the low energy band is relevant and that the model could be well approximated by some effective one-band bare band coupled to phonons. If that was true, its ARPES intensity would allow one to extract a selfenergy with the methods employed in one-band systems. 

 Our main result is that such assumptions may fail because even though the two bare bands are well separated in energy (on the scale of the EPC), they consist of linear combinations of the same underlying valence orbitals. The relative character ($s$ vs. $p$) of the lower band changes as function of the momentum. Combined with the fact that different orbitals have different matrix elements, we conclude that the accurate extraction of the selfenergy matrix from the ARPES intensity becomes essentially impossible  in such cases. Of course, one can still apply the  ARPES analysis to extract a 'selfenergy' -- our point is that this quantity is not related in any simple way to the theoretically calculated selfenergy matrix.

 We believe that this conclusion is very important, considering how ubiquitous the use of ARPES data has become in characterizing complex quantum materials.

 Our second result is the demonstration that having constant energy ARPES cuts with a Lorentzian peak is not necessarily correlated with a local selfenergy. If the selfenergy happens to be local, one can indeed extract its energy dependence from the usual ARPES analysis, as we showed for the Holstein EPC. However, like shown for the Peierls case, it is possible to have a momentum-dependent selfenergy which produces Lorentzian peaks in the constant energy cuts. The usual ARPES analysis will now extract a wrong, local selfenergy. 

 Our overall conclusion is that a lot of care and significant theoretical support needs to be used when interpreting ARPES data for complex materials, otherwise one might arrive at overly simplified (if not completely wrong) conclusions. Clearly, a lot of work remains to be done to fully understand these issues.

\acknowledgements
We thank Andrea Damascelli and Jorg Fink for comments and suggestions. We acknowledge support from the Natural Sciences and Engineering Research Council of Canada (NSERC), the Stewart
Blusson Quantum Matter Institute (SBQMI) and the
Max-Planck-UBC-UTokyo Center for Quantum Materials.

\bibliography{interband_ref}

\begin{thebibliography}{21}%
\makeatletter
\providecommand \@ifxundefined [1]{%
 \@ifx{#1\undefined}
}%
\providecommand \@ifnum [1]{%
 \ifnum #1\expandafter \@firstoftwo
 \else \expandafter \@secondoftwo
 \fi
}%
\providecommand \@ifx [1]{%
 \ifx #1\expandafter \@firstoftwo
 \else \expandafter \@secondoftwo
 \fi
}%
\providecommand \natexlab [1]{#1}%
\providecommand \enquote  [1]{``#1''}%
\providecommand \bibnamefont  [1]{#1}%
\providecommand \bibfnamefont [1]{#1}%
\providecommand \citenamefont [1]{#1}%
\providecommand \href@noop [0]{\@secondoftwo}%
\providecommand \href [0]{\begingroup \@sanitize@url \@href}%
\providecommand \@href[1]{\@@startlink{#1}\@@href}%
\providecommand \@@href[1]{\endgroup#1\@@endlink}%
\providecommand \@sanitize@url [0]{\catcode `\\12\catcode `\$12\catcode
  `\&12\catcode `\#12\catcode `\^12\catcode `\_12\catcode `\%12\relax}%
\providecommand \@@startlink[1]{}%
\providecommand \@@endlink[0]{}%
\providecommand \url  [0]{\begingroup\@sanitize@url \@url }%
\providecommand \@url [1]{\endgroup\@href {#1}{\urlprefix }}%
\providecommand \urlprefix  [0]{URL }%
\providecommand \Eprint [0]{\href }%
\providecommand \doibase [0]{http://dx.doi.org/}%
\providecommand \selectlanguage [0]{\@gobble}%
\providecommand \bibinfo  [0]{\@secondoftwo}%
\providecommand \bibfield  [0]{\@secondoftwo}%
\providecommand \translation [1]{[#1]}%
\providecommand \BibitemOpen [0]{}%
\providecommand \bibitemStop [0]{}%
\providecommand \bibitemNoStop [0]{.\EOS\space}%
\providecommand \EOS [0]{\spacefactor3000\relax}%
\providecommand \BibitemShut  [1]{\csname bibitem#1\endcsname}%
\let\auto@bib@innerbib\@empty
\bibitem [{\citenamefont {Damascelli}\ \emph {et~al.}(2003)\citenamefont
  {Damascelli}, \citenamefont {Hussain},\ and\ \citenamefont {Shen}}]{Dam2003}%
  \BibitemOpen
  \bibfield  {author} {\bibinfo {author} {\bibfnamefont {A.}~\bibnamefont
  {Damascelli}}, \bibinfo {author} {\bibfnamefont {Z.}~\bibnamefont {Hussain}},
  \ and\ \bibinfo {author} {\bibfnamefont {Z.-X.}\ \bibnamefont {Shen}},\
  }\href {\doibase 10.1103/RevModPhys.75.473} {\bibfield  {journal} {\bibinfo
  {journal} {Rev. Mod. Phys.}\ }\textbf {\bibinfo {volume} {75}},\ \bibinfo
  {pages} {473} (\bibinfo {year} {2003})}\BibitemShut {NoStop}%
\bibitem [{\citenamefont {Lau}\ \emph {et~al.}(2007)\citenamefont {Lau},
  \citenamefont {Berciu},\ and\ \citenamefont {Sawatzky}}]{BLau2007}%
  \BibitemOpen
  \bibfield  {author} {\bibinfo {author} {\bibfnamefont {B.}~\bibnamefont
  {Lau}}, \bibinfo {author} {\bibfnamefont {M.}~\bibnamefont {Berciu}}, \ and\
  \bibinfo {author} {\bibfnamefont {G.~A.}\ \bibnamefont {Sawatzky}},\ }\href
  {\doibase 10.1103/PhysRevB.76.174305} {\bibfield  {journal} {\bibinfo
  {journal} {Phys. Rev. B}\ }\textbf {\bibinfo {volume} {76}},\ \bibinfo
  {pages} {174305} (\bibinfo {year} {2007})}\BibitemShut {NoStop}%
\bibitem [{\citenamefont {Holstein}(1959)}]{HolsteinI}%
  \BibitemOpen
  \bibfield  {author} {\bibinfo {author} {\bibfnamefont {T.}~\bibnamefont
  {Holstein}},\ }\href {\doibase doi.org/10.1016/0003-4916(59)90002-8}
  {\bibfield  {journal} {\bibinfo  {journal} {Ann. Phys. (N.Y.)}\ }\textbf
  {\bibinfo {volume} {8}},\ \bibinfo {pages} {325} (\bibinfo {year}
  {1959})}\BibitemShut {NoStop}%
\bibitem [{\citenamefont {Goodvin}\ and\ \citenamefont
  {Berciu}(2008)}]{GGoodvinPRB2008}%
  \BibitemOpen
  \bibfield  {author} {\bibinfo {author} {\bibfnamefont {G.~L.}\ \bibnamefont
  {Goodvin}}\ and\ \bibinfo {author} {\bibfnamefont {M.}~\bibnamefont
  {Berciu}},\ }\href {\doibase 10.1103/PhysRevB.78.235120} {\bibfield
  {journal} {\bibinfo  {journal} {Phys. Rev. B}\ }\textbf {\bibinfo {volume}
  {78}},\ \bibinfo {pages} {235120} (\bibinfo {year} {2008})}\BibitemShut
  {NoStop}%
\bibitem [{\citenamefont {Marchand}\ \emph {et~al.}(2017)\citenamefont
  {Marchand}, \citenamefont {Stamp},\ and\ \citenamefont
  {Berciu}}]{DMarchandPRB2017}%
  \BibitemOpen
  \bibfield  {author} {\bibinfo {author} {\bibfnamefont {D.~J.~J.}\
  \bibnamefont {Marchand}}, \bibinfo {author} {\bibfnamefont {P.~C.~E.}\
  \bibnamefont {Stamp}}, \ and\ \bibinfo {author} {\bibfnamefont
  {M.}~\bibnamefont {Berciu}},\ }\href {\doibase 10.1103/PhysRevB.95.035117}
  {\bibfield  {journal} {\bibinfo  {journal} {Phys. Rev. B}\ }\textbf {\bibinfo
  {volume} {95}},\ \bibinfo {pages} {035117} (\bibinfo {year}
  {2017})}\BibitemShut {NoStop}%
\bibitem [{\citenamefont {Herrera}\ \emph {et~al.}(2013)\citenamefont
  {Herrera}, \citenamefont {Madison}, \citenamefont {Krems},\ and\
  \citenamefont {Berciu}}]{HerreraPRL2013}%
  \BibitemOpen
  \bibfield  {author} {\bibinfo {author} {\bibfnamefont {F.}~\bibnamefont
  {Herrera}}, \bibinfo {author} {\bibfnamefont {K.~W.}\ \bibnamefont
  {Madison}}, \bibinfo {author} {\bibfnamefont {R.~V.}\ \bibnamefont {Krems}},
  \ and\ \bibinfo {author} {\bibfnamefont {M.}~\bibnamefont {Berciu}},\ }\href
  {\doibase 10.1103/PhysRevLett.110.223002} {\bibfield  {journal} {\bibinfo
  {journal} {Phys. Rev. Lett.}\ }\textbf {\bibinfo {volume} {110}},\ \bibinfo
  {pages} {223002} (\bibinfo {year} {2013})}\BibitemShut {NoStop}%
\bibitem [{\citenamefont {M\"oller}\ and\ \citenamefont
  {Berciu}(2016)}]{Mol2016}%
  \BibitemOpen
  \bibfield  {author} {\bibinfo {author} {\bibfnamefont {M.~M.}\ \bibnamefont
  {M\"oller}}\ and\ \bibinfo {author} {\bibfnamefont {M.}~\bibnamefont
  {Berciu}},\ }\href {\doibase 10.1103/PhysRevB.93.035130} {\bibfield
  {journal} {\bibinfo  {journal} {Phys. Rev. B}\ }\textbf {\bibinfo {volume}
  {93}},\ \bibinfo {pages} {035130} (\bibinfo {year} {2016})}\BibitemShut
  {NoStop}%
\bibitem [{\citenamefont {M{\"o}ller}\ \emph {et~al.}(2017)\citenamefont
  {M{\"o}ller}, \citenamefont {Sawatzky}, \citenamefont {Franz},\ and\
  \citenamefont {Berciu}}]{Mol2017}%
  \BibitemOpen
  \bibfield  {author} {\bibinfo {author} {\bibfnamefont {M.~M.}\ \bibnamefont
  {M{\"o}ller}}, \bibinfo {author} {\bibfnamefont {G.~A.}\ \bibnamefont
  {Sawatzky}}, \bibinfo {author} {\bibfnamefont {M.}~\bibnamefont {Franz}}, \
  and\ \bibinfo {author} {\bibfnamefont {M.}~\bibnamefont {Berciu}},\ }\href
  {\doibase 10.1038/s41467-017-02442-y} {\bibfield  {journal} {\bibinfo
  {journal} {Nature Communications}\ }\textbf {\bibinfo {volume} {8}},\
  \bibinfo {pages} {2267} (\bibinfo {year} {2017})}\BibitemShut {NoStop}%
\bibitem [{\citenamefont {Yam}\ \emph {et~al.}(2020)\citenamefont {Yam},
  \citenamefont {Moeller}, \citenamefont {Sawatzky},\ and\ \citenamefont
  {Berciu}}]{Yam2020}%
  \BibitemOpen
  \bibfield  {author} {\bibinfo {author} {\bibfnamefont {Y.-C.}\ \bibnamefont
  {Yam}}, \bibinfo {author} {\bibfnamefont {M.~M.}\ \bibnamefont {Moeller}},
  \bibinfo {author} {\bibfnamefont {G.~A.}\ \bibnamefont {Sawatzky}}, \ and\
  \bibinfo {author} {\bibfnamefont {M.}~\bibnamefont {Berciu}},\ }\href
  {\doibase 10.1103/PhysRevB.102.235145} {\bibfield  {journal} {\bibinfo
  {journal} {Phys. Rev. B}\ }\textbf {\bibinfo {volume} {102}},\ \bibinfo
  {pages} {235145} (\bibinfo {year} {2020})}\BibitemShut {NoStop}%
\bibitem [{\citenamefont {Berciu}(2006)}]{Ber2006}%
  \BibitemOpen
  \bibfield  {author} {\bibinfo {author} {\bibfnamefont {M.}~\bibnamefont
  {Berciu}},\ }\href {\doibase 10.1103/PhysRevLett.97.036402} {\bibfield
  {journal} {\bibinfo  {journal} {Phys. Rev. Lett.}\ }\textbf {\bibinfo
  {volume} {97}},\ \bibinfo {pages} {036402} (\bibinfo {year}
  {2006})}\BibitemShut {NoStop}%
\bibitem [{\citenamefont {Berciu}\ and\ \citenamefont
  {Goodvin}(2007)}]{Ber2007}%
  \BibitemOpen
  \bibfield  {author} {\bibinfo {author} {\bibfnamefont {M.}~\bibnamefont
  {Berciu}}\ and\ \bibinfo {author} {\bibfnamefont {G.~L.}\ \bibnamefont
  {Goodvin}},\ }\href {\doibase 10.1103/PhysRevB.76.165109} {\bibfield
  {journal} {\bibinfo  {journal} {Phys. Rev. B}\ }\textbf {\bibinfo {volume}
  {76}},\ \bibinfo {pages} {165109} (\bibinfo {year} {2007})}\BibitemShut
  {NoStop}%
\bibitem [{\citenamefont {Varma}(1988)}]{Var1988}%
  \BibitemOpen
  \bibfield  {author} {\bibinfo {author} {\bibfnamefont {C.~M.}\ \bibnamefont
  {Varma}},\ }\href {\doibase 10.1103/PhysRevLett.61.2713} {\bibfield
  {journal} {\bibinfo  {journal} {Phys. Rev. Lett.}\ }\textbf {\bibinfo
  {volume} {61}},\ \bibinfo {pages} {2713} (\bibinfo {year}
  {1988})}\BibitemShut {NoStop}%
\bibitem [{\citenamefont {Hase}\ and\ \citenamefont
  {Yanagisawa}(2007)}]{Has2007}%
  \BibitemOpen
  \bibfield  {author} {\bibinfo {author} {\bibfnamefont {I.}~\bibnamefont
  {Hase}}\ and\ \bibinfo {author} {\bibfnamefont {T.}~\bibnamefont
  {Yanagisawa}},\ }\href {\doibase 10.1103/PhysRevB.76.174103} {\bibfield
  {journal} {\bibinfo  {journal} {Phys. Rev. B}\ }\textbf {\bibinfo {volume}
  {76}},\ \bibinfo {pages} {174103} (\bibinfo {year} {2007})}\BibitemShut
  {NoStop}%
\bibitem [{\citenamefont {Khazraie}\ \emph {et~al.}(2018)\citenamefont
  {Khazraie}, \citenamefont {Foyevtsova}, \citenamefont {Elfimov},\ and\
  \citenamefont {Sawatzky}}]{Kha2018}%
  \BibitemOpen
  \bibfield  {author} {\bibinfo {author} {\bibfnamefont {A.}~\bibnamefont
  {Khazraie}}, \bibinfo {author} {\bibfnamefont {K.}~\bibnamefont
  {Foyevtsova}}, \bibinfo {author} {\bibfnamefont {I.}~\bibnamefont {Elfimov}},
  \ and\ \bibinfo {author} {\bibfnamefont {G.~A.}\ \bibnamefont {Sawatzky}},\
  }\href {\doibase 10.1103/PhysRevB.97.075103} {\bibfield  {journal} {\bibinfo
  {journal} {Phys. Rev. B}\ }\textbf {\bibinfo {volume} {97}},\ \bibinfo
  {pages} {075103} (\bibinfo {year} {2018})}\BibitemShut {NoStop}%
\bibitem [{\citenamefont {Ebrahimnejad}\ \emph {et~al.}(2014)\citenamefont
  {Ebrahimnejad}, \citenamefont {Sawatzky},\ and\ \citenamefont
  {Berciu}}]{Ebr2014}%
  \BibitemOpen
  \bibfield  {author} {\bibinfo {author} {\bibfnamefont {H.}~\bibnamefont
  {Ebrahimnejad}}, \bibinfo {author} {\bibfnamefont {G.~A.}\ \bibnamefont
  {Sawatzky}}, \ and\ \bibinfo {author} {\bibfnamefont {M.}~\bibnamefont
  {Berciu}},\ }\href {\doibase 10.1038/nphys3130} {\bibfield  {journal}
  {\bibinfo  {journal} {Nature Physics}\ }\textbf {\bibinfo {volume} {10}},\
  \bibinfo {pages} {951} (\bibinfo {year} {2014})}\BibitemShut {NoStop}%
\bibitem [{\citenamefont {Veenstra}(2007)}]{Nei2007}%
  \BibitemOpen
  \bibfield  {author} {\bibinfo {author} {\bibfnamefont {C.~N.}\ \bibnamefont
  {Veenstra}},\ }\emph {\bibinfo {title} {Spectral function analysis on the
  Holstein polaron problem: extraction of the self-energy and coupling
  strength, and their implications for angle resolved photoemission
  spectroscopy}},\ \href {\doibase http://dx.doi.org/10.14288/1.0084909} {Ph.D.
  thesis},\ \bibinfo  {school} {University of British Columbia} (\bibinfo
  {year} {2007})\BibitemShut {NoStop}%
\bibitem [{\citenamefont {Veenstra}\ \emph {et~al.}(2010)\citenamefont
  {Veenstra}, \citenamefont {Goodvin}, \citenamefont {Berciu},\ and\
  \citenamefont {Damascelli}}]{Vee10}%
  \BibitemOpen
  \bibfield  {author} {\bibinfo {author} {\bibfnamefont {C.~N.}\ \bibnamefont
  {Veenstra}}, \bibinfo {author} {\bibfnamefont {G.~L.}\ \bibnamefont
  {Goodvin}}, \bibinfo {author} {\bibfnamefont {M.}~\bibnamefont {Berciu}}, \
  and\ \bibinfo {author} {\bibfnamefont {A.}~\bibnamefont {Damascelli}},\
  }\href {\doibase 10.1103/PhysRevB.82.012504} {\bibfield  {journal} {\bibinfo
  {journal} {Phys. Rev. B}\ }\textbf {\bibinfo {volume} {82}},\ \bibinfo
  {pages} {012504} (\bibinfo {year} {2010})}\BibitemShut {NoStop}%
\bibitem [{\citenamefont {Veenstra}\ \emph {et~al.}(2011)\citenamefont
  {Veenstra}, \citenamefont {Goodvin}, \citenamefont {Berciu},\ and\
  \citenamefont {Damascelli}}]{Vee11}%
  \BibitemOpen
  \bibfield  {author} {\bibinfo {author} {\bibfnamefont {C.~N.}\ \bibnamefont
  {Veenstra}}, \bibinfo {author} {\bibfnamefont {G.~L.}\ \bibnamefont
  {Goodvin}}, \bibinfo {author} {\bibfnamefont {M.}~\bibnamefont {Berciu}}, \
  and\ \bibinfo {author} {\bibfnamefont {A.}~\bibnamefont {Damascelli}},\
  }\href {\doibase 10.1103/PhysRevB.84.085126} {\bibfield  {journal} {\bibinfo
  {journal} {Phys. Rev. B}\ }\textbf {\bibinfo {volume} {84}},\ \bibinfo
  {pages} {085126} (\bibinfo {year} {2011})}\BibitemShut {NoStop}%
\bibitem [{\citenamefont {Day}\ \emph {et~al.}(2019)\citenamefont {Day},
  \citenamefont {Zwartsenberg}, \citenamefont {Elfimov},\ and\ \citenamefont
  {Damascelli}}]{Day2019}%
  \BibitemOpen
  \bibfield  {author} {\bibinfo {author} {\bibfnamefont {R.~P.}\ \bibnamefont
  {Day}}, \bibinfo {author} {\bibfnamefont {B.}~\bibnamefont {Zwartsenberg}},
  \bibinfo {author} {\bibfnamefont {I.~S.}\ \bibnamefont {Elfimov}}, \ and\
  \bibinfo {author} {\bibfnamefont {A.}~\bibnamefont {Damascelli}},\ }\href
  {\doibase 10.1038/s41535-019-0194-8} {\bibfield  {journal} {\bibinfo
  {journal} {npj Quantum Materials}\ }\textbf {\bibinfo {volume} {4}},\
  \bibinfo {pages} {54} (\bibinfo {year} {2019})}\BibitemShut {NoStop}%
\bibitem [{\citenamefont {Day}(2020)}]{Day2020}%
  \BibitemOpen
  \bibfield  {author} {\bibinfo {author} {\bibfnamefont {R.~P.}\ \bibnamefont
  {Day}},\ }\emph {\bibinfo {title} {Leveraging the light-matter interaction in
  angle-resolved photoemission spectroscopy}},\ \href {\doibase
  http://dx.doi.org/10.14288/1.0392706} {Ph.D. thesis},\ \bibinfo  {school}
  {University of British Columbia} (\bibinfo {year} {2020})\BibitemShut
  {NoStop}%
\bibitem [{\citenamefont {Derriche}\ \emph {et~al.}(2022)\citenamefont
  {Derriche}, \citenamefont {Elfimov},\ and\ \citenamefont
  {Sawatzky}}]{Der2022}%
  \BibitemOpen
  \bibfield  {author} {\bibinfo {author} {\bibfnamefont {N.}~\bibnamefont
  {Derriche}}, \bibinfo {author} {\bibfnamefont {I.}~\bibnamefont {Elfimov}}, \
  and\ \bibinfo {author} {\bibfnamefont {G.}~\bibnamefont {Sawatzky}},\ }\href
  {\doibase 10.1103/PhysRevB.106.064102} {\bibfield  {journal} {\bibinfo
  {journal} {Phys. Rev. B}\ }\textbf {\bibinfo {volume} {106}},\ \bibinfo
  {pages} {064102} (\bibinfo {year} {2022})}\BibitemShut {NoStop}%
\end{thebibliography}%

\end{document}